\documentclass[
,twocolumn
]{aastex631}

\usepackage{amsmath, subfigure}
\usepackage{threeparttable}
\usepackage{orcidlink}
\usepackage{url}

\begin{document}
\begin{sloppypar}

\title{Coherence of Supermassive Black Hole Binary Demographics with the nHz Stochastic Gravitational Wave Background}

\author{Katsunori Kusakabe}
\affiliation{Department of Earth and Space Science, Graduate School of Science, The University of Osaka, 1-1 Machikaneyama, Toyonaka, Osaka 560-0043, Japan}

\author[0000-0002-7272-1136]{Yoshiyuki Inoue}    
\affiliation{Department of Earth and Space Science, Graduate School of Science, The University of Osaka, 1-1 Machikaneyama, Toyonaka, Osaka 560-0043, Japan}
\affiliation{Interdisciplinary Theoretical \& Mathematical Science Center (iTHEMS), RIKEN, 2-1 Hirosawa, 351-0198, Japan}
\affiliation{Kavli Institute for the Physics and Mathematics of the Universe (WPI), UTIAS, The University of Tokyo, 5-1-5 Kashiwanoha, Kashiwa, Chiba 277-8583, Japan}

\author[0000-0003-3467-6079]{Daisuke Toyouchi}
\affiliation{Department of Earth and Space Science, Graduate School of Science, The University of Osaka, 1-1 Machikaneyama, Toyonaka, Osaka 560-0043, Japan}

\correspondingauthor{Katsunori Kusakabe}
\email{u778278e@ecs.osaka-u.ac.jp}

\begin{abstract}
We present a refined estimation of the stochastic gravitational wave background (SGWB) based on observed dual active galactic nuclei (AGNs) together with AGN X-ray luminosity functions, in light of recent Pulsar Timing Array detections of an nHz SGWB. We identify a characteristic luminosity dependence in dual AGN fractions by compiling recent observational datasets, providing crucial constraints on supermassive black hole binary (SMBHB) populations. Our AGN-based model reproduces the current SGWB measurements within PTA observational uncertainties of $2\text{--}4\ \sigma$, demonstrating consistency between electromagnetic and gravitational wave observations.  These findings establish SMBHBs as the dominant source of the nHz gravitational wave signal, providing crucial insights into their demographics and evolution.
\end{abstract}

\keywords{Gravitational waves, Gravitational wave sources, Supermassive black holes, Active galactic nuclei, Double quasars}

% %%%%%%%%%%%%%%%%%%%%%%%%%%%%%%%%%%%%%%%%%%%
\section{Introduction}\label{sec:intro} 
% %%%%%%%%%%%%%%%%%%%%%%%%%%%%%%%%%%%%%%%%%%%
Recent detections of a stochastic gravitational wave background (SGWB) in the nanohertz frequency range by Pulsar Timing Array (PTA) collaborations \citep{NANOGrav:2023hde, NANOGrav:2023gor, EPTA:2023fyk, EPTA:2023xxk, Reardon:2023zen, Reardon:2023gzh, Xu:2023wog} have reignited interest in its potential sources. While various cosmological origins have been proposed \citep{Vilenkin:1984ib, Hindmarsh:1994re, Caprini:2015zlo, Guzzetti:2016mkm, Saikawa:2017hiv, Caprini:2019egz, Hindmarsh:2020hop, Domenech:2021ztg, Yuan:2021qgz, NANOGrav:2023hvm}, inspiraling supermassive black hole binaries (SMBHBs) remain the primary candidates \citep{Rajagopal:1994zj, Jaffe:2002rt, Wyithe:2002ep, Sesana:2004sp, Enoki:2004ew, Enoki:2006kj, Burke-Spolaor:2018bvk}.

Galaxy merger models incorporating SMBHB dynamics \citep[e.g.,][]{Jaffe:2002rt, Sesana:2008xk, 2011MNRAS.411.1467K, Sesana:2012ak, Sesana:2013wja, Chen:2018znx, Chen:2020qlp, Bi:2023tib} have provided insights into their SGWB contribution. However, the detected signal amplitudes exceed predictions in some models, implying that supermassive black holes (SMBHs) with $10^9 M_{\odot}$ are more abundant than empirically expected \citep{Sato-Polito:2023gym, Sato-Polito:2024lew, 2025arXiv250201024C, Lapi:2025wxt, Sato-Polito:2025dqw}. 

This discrepancy highlights two key uncertainties in galaxy merger models. The first key uncertainty lies in estimating SMBH masses, particularly given by recent observations suggesting redshift-dependent variations in galaxy-SMBH mass correlations \citep{2019PASJ...71..111I, 2020A&A...637A..84P, 2021ApJ...914...36I, 2023ApJ...959...39H, 2024A&A...691A.145M, 2024ApJ...966L..30M}. The SMBH mass function (BHMF) can be alternatively derived through active galactic nucleus (AGN) luminosity functions, so-called the Soltan argument \citep{Soltan:1982vf, 1992MNRAS.259..725S, Yu:2002sq}, which have successfully reproduced local observations \citep{2012AdAst2012E...7K, Ueda:2014tma, Tucci:2016tyc, Shen:2020obl}.

Another key uncertainty arises from the pair formation rate of SMBHs. Galaxy merger models typically assume instantaneous SMBHB formation after galaxy merger \citep{Burke-Spolaor:2018bvk}, resulting in coalescence in a certain timescale \citep{Begelman:1980vb}, with merger rates derived from cosmological simulations \citep[e.g.,][]{Ravi:2014aha, Kelley:2016gse, Kelley:2017lek} or semi-analytical models (SAMs) \citep[e.g.,][]{2011MNRAS.411.1467K, Sesana:2012ak, Chen:2020qlp}. However, these predictions can vary significantly with assumed merger timescales, environmental effects, and binary formation efficiency (see e.g., \citet{2019astro2020T.504K} and references therein). 

A quasar-based SMBHB population model has also been developed recently in the literature, utilizing quasar luminosity functions to construct a BH-observable dependent model \citep{2009ApJ...700.1952H, Goulding:2019hnn, Xin:2021mmk, Casey-Clyde:2021xro, Kis-Toth:2024gkm, Xin:2025voy, Lapi:2025wxt}. These models explore various connections between quasar activity and SMBHB populations. For instance, \cite{Casey-Clyde:2021xro} develops a SAM linking quasar populations to the SGWB, exploring the fraction of SMBHBs associated with quasars. However, significant uncertainties persist concerning the precise rate at which quasars form as binaries or their true binary occupation fraction. 

Recently, direct observational constraints on SMBHB merger rates have been provided by dual AGN studies through various approaches: orbital motion measurements \citep{Sudou:2003hv, Bon:2016jtk, Runnoe:2017oxn,Guo:2018xum,Adhikari:2023uju}, spectroscopic separation \citep{Zhou:2004uq, Comerford:2008gm, 2012ApJ...759..118B, Comerford:2013fha, 2014ApJ...789..140L}, and high-resolution imaging \citep{Rodriguez:2006th,Deane:2014jqa, Comerford:2015qda, 2017NatAs...1..727K}. Furthermore, very recent observations, including those from JWST, have extended this exploration to wide luminosity ranges and high redshifts \citep{Goulding:2023gqa,2023arXiv231003067P,2024MNRAS.531..355U, Kovacs:2024zfh}, providing implications for dual AGN activity \citep{Padmanabhan:2024nvv} and the interactions of SMBHBs with their surrounding environments \citep{Ellis:2024wdh}. These studies constrain the dual AGN fraction ($f_\mathrm{dual}$) between $\sim0.01$\% and $\sim20$\% \citep{2011ApJ...737..101L, 2012ApJ...746L..22K, 2020ApJ...899..154S, Shen:2022cmp, 2023arXiv231003067P,2024arXiv240514980L}. These dual AGN observations enable us to probe SMBHB populations directly.

In this paper, we compile recent measurements of the dual AGN fraction and use them together with the AGN X-ray luminosity function to construct the SMBHB mass function, thereby constraining their demographics and merger rates. Based on this AGN-based population model, we further estimate the expected contribution of SMBHBs to the SGWB. Our methodology for computing the SGWB is described in Section \ref{sec:Model}, followed by the description and compilation of the dual AGN fraction in Section \ref{sec:Dual_AGN}. Section \ref{sec:SGWB} presents the results of the expected SGWB from SMBHBs. Subsequent discussions, including the possible parameter impact and the comparison with previous works, are given in Section \ref{sec:Discussion}, and conclusions are in Section \ref{sec:Conclusion}. We adopt standard cosmological parameters with $(H_0, \Omega_m, \Omega_{\Lambda})= (67.4~ \mathrm{km}~ \mathrm{s}^{-1}~ \mathrm{Mpc}^{-1}, 0.315, 0.685)$ \citep{Planck:2018vyg}.

% %%%%%%%%%%%%%%%%%%%%%%%%%%%%%%%%%%%%%%%%%%%
\section{AGN-based SMBHB Population Synthesis}\label{sec:Model}
% %%%%%%%%%%%%%%%%%%%%%%%%%%%%%%%%%%%%%%%%%%%
We estimate the SGWB by integrating gravitational waves emitted by SMBHBs over cosmic history. The energy density of the SGWB $\Omega_{\mathrm{GW}}(f)$ at observed frequency $f$ is given by \citep{2001astro.ph..8028P}:
\begin{equation}
\Omega_{\mathrm{GW}} (f) = \frac{8\pi Gf}{3 {H_0}^2 c^2} \int dz d\mathcal{M} \frac{d^2 n}{dz d\mathcal{M}} \frac{dE_\mathrm{GW}}{d f_r},
\label{eq:Phinney}
\end{equation}
where $G$ is the gravitational constant, $c$ is the speed of light, $z$ is redshift, $\mathcal{M} =Mq^{\frac{3}{5}}/(1+q)^{\frac{1}{5}}$ is the chirp mass, $M$ is the primary SMBH mass, $q$ is the mass ratio, $n$ is the comoving number density of sources with $d^2 n/dz d\mathcal{M}$ represents the SMBHB merger rate, and $dE_{\rm GW}/df_r$ is the energy spectrum with $f_r = (1+z)f$. 

We adopt an energy spectrum reproducing the inspiral phase with the initial eccentricity \citep{Enoki:2006kj, Chen:2016zyo, Bi:2023tib}. In the framework of our SMBHB orbital prescription, including gas-rich environments, the binary evolution transitions from gas-driven migration to GW-driven inspiral. We define the spectral turnover frequency, $f_t$, as the frequency where the evolutionary rates of these two mechanisms become comparable, i.e., $(df/dt)_{\rm gas} \simeq (df/dt)_{\rm GW}$. This transition creates a characteristic break in the GW spectrum. The detailed derivation of $f_t$ and the specific evolutionary timescales are presented in the descriptions below. At this transition epoch, we set the initial eccentricity at the beginning of the GW emission as $e_0 = 0.45$, which is implied by the results of recent suites of hydrodynamical simulations \citep{Munoz:2018tnj, Zrake:2020zkw, DOrazio:2021kob, Siwek:2023rlk}. The uncertainties in the initial eccentricity primarily affect the GW spectrum in the very low-frequency regime \citep{Sesana:2013wja, Ravi:2014aha, Kelley:2017lek, Moreschi:2025qtm, Sah:2025dmv}. We discuss the impact and implications of eccentricity in Sec. \ref{subsec:eccentricity}. 

We estimate the SMBHB merger rate by combining the BHMF $\Phi_{\mathrm{BH}} (M,z)$, merger timescale $\tau (M,q)$ and dual AGN fraction $f_\mathrm{dual}(M)$ (see Sec.~\ref{sec:Dual_AGN} for a detailed description of $f_\mathrm{dual}$) as analogs of galaxy evolution models \citep{Sesana:2008xk, Sesana:2013wja, Chen:2018znx, Bi:2023tib}. Here, we define the dual AGN fraction as the ratio of the number density 
of dual AGN systems to the total number density of AGN. Given the weak redshift dependence of $f_\mathrm{dual}(M)$ \citep{2020ApJ...899..154S,Shen:2022cmp,2024arXiv240514980L}, we adopt redshift independent form. We consider a uniform distribution for the mass ratio $q$ between 0.1 and 1 since the lower limit of the flux ratio in the dual AGN observations is approximately $10:1$ \citep{2020ApJ...899..154S, Shen:2022cmp, 2024arXiv240514980L}. We note that cosmological simulations such as Illustris predict a non-uniform mass ratio distribution that depends on the total binary mass \citep{2016MNRAS.458.1013S, Kelley:2016gse, 2022MNRAS.514..640V, Saeedzadeh:2024wkz}. However, given the uncertainties in the observational constraints on the mass ratio distribution, we adopt a uniform distribution for simplicity. As for the range of mass raito, while binaries with highly unequal mass ratios may exist and could be missed by current observations, their contribution to the overall SGWB is limited due to the strong dependence of the GW energy spectrum on chirp mass (e.g., the GW emission with a mass ratio of $q=0.01$ is about 100 times weaker than that from an equal-mass binary with the same primary BH mass). The integrand in Eq.~\ref{eq:Phinney} can be expressed as

\begin{align}
\frac{d^2 n}{dz d\mathcal{M}}\frac{dE_\mathrm{GW}}{d f_r} &= \int \frac{\Phi_\mathrm{BH}(M,z)}{M \mathrm{ln}10} \mathcal{R}_{\mathrm{eff}} (M,q,z) \nonumber \\
&\times \frac{dE_\mathrm{GW} (\mathcal{M},z,q,f)}{d f_r}\frac{d M}{d\mathcal{M}}dq.
\end{align}
Here, $\mathcal{R}_{\mathrm{eff}}(M, q, z)$ is a term to account for both the depletion of SMBH populations and mass growth through mergers. This effective correction term $\mathcal{R}_{\mathrm{eff}}(M, q, z)$ is given as
\begin{flalign}
& \mathcal{R}_{\mathrm{eff}}(M,q,z) = && \nonumber \\
& \quad
\begin{cases}
\mathcal{N}(z) & \hspace{-8mm} (\mathcal{N} < 1), \\
\begin{aligned}
\int_{0}^{\mathcal{N}(z)} dk& \Bigg( \frac{1}{2^{k}} \frac{f_{\mathrm{dual}}\left[(1+q)^{k} M\right]}{\tau\left[(1+q)^{k} M,q\right]} \\
& \Big/ \frac{f_{\mathrm{dual}}(M)}{\tau(M,q)} \Bigg) S(k,q) \\ 
\times \Theta&(t_{\mathrm{age}}(z) - \tau\left[(1+q)^{k} M,q\right])) 
\end{aligned}
& \hspace{-8mm} (\mathcal{N} \ge 1),
\end{cases} &
\end{flalign}
where $\mathcal{N} (z) \equiv ({f_\mathrm{dual}}/{\tau}) \cdot (dt / dz)$ represents the number of merger events per BH at redshift $z$ and $k$ is the integral variable counting the cumulative number of merger events. When $\mathcal{N}\ge1$, this term tracks the reduction in black hole numbers ($1/2^{k}$ factor), mass growth through mergers, and GW emission efficiency changes through the function $S(k,q)$ defined as \citep{Sato-Polito:2023gym} 
\begin{equation}
    S(k,q) = 4\zeta \sum_{k^{\prime}=0}^k\left(\frac{1+q^{5/3}}{(1+q)^{5/3}} \right)^{k^{\prime}}= 4\zeta \frac{1-\left(\frac{1+q^{5/3}}{(1+q)^{5/3}} \right)^{k+1}}{1- \left(\frac{1+q^{5/3}}{(1+q)^{5/3}} \right)},
    \label{eq:GW_efficiency}
\end{equation}
where $\zeta \equiv q/(1+q)^2$ is the symmetric mass ratio that characterizes the GW emission efficiency. The summation variable $k^{\prime}$ in $S(k,q)$ tracks the GW efficiency changes at each merger step. The Heaviside step function, $\Theta$, serves as a temporal filter, ensuring that only SMBHB merger events completed by the cosmic age at observed redshift $z$ given as $t_{\mathrm{age}}(z) = \int_{z}^{\infty} {dz'}/{H_0}{(1+z')\sqrt{\Omega_m(1+z')^3 + \Omega_\Lambda}}$. This reflects our model framework, where all SMBHBs are initialized at an initial separation and evolve according to the merger timescale prescription (see below for merger timescale prescription). Note that the ratio $f_{\rm dual}/\tau$ in $\mathcal{R}_{\rm eff}$ effectively converts the observed dual AGN fraction into a merger rate: while long-lived binaries have higher detection probabilities (larger $f_{\rm dual}$), dividing by their correspondingly longer merger timescales $\tau$ yields the correct rate of coalescence events per unit time, thereby avoiding overcounting of persistent systems across redshift intervals. These comprehensive treatments prevent the overestimation of SMBH merger rates by accounting for the reduction of the merging SMBH population, and systematic shifts in binary parameters such as merger timescales and dual AGN fractions through successive mergers, in addition to the enhanced GW emission described by $S(k,q)$. Thereby, the term $\mathcal{R}_{\mathrm{eff}}$ provides a framework for capturing the hierarchical growth of SMBHs.

In the BH population analysis, we utilize the BHMF derived from the AGN X-ray luminosity function and Eddington ratio distribution. The preference for X-ray LFs stems from their robustness against dust obscuration, offering a more complete census of intrinsic AGN activity \citep{Ueda:2014tma, Aird:2015fya}. The AGN mass function  $\Psi_{\mathrm{AGN}}$ is expressed as
\begin{equation}
\Psi_\mathrm{AGN}(M,z) = \frac{d \log M}{dM} \int \frac{d \Psi_\mathrm{bol} (L_\mathrm{bol},z)}{ d \log L_\mathrm{bol}} P(\lambda | L_\mathrm{bol} ,z) d\log \lambda,
\end{equation}
where $d \Psi_\mathrm{bol} (L_\mathrm{bol},z)/ d \log L_\mathrm{bol}$ is the AGN bolometric luminosity function estimated from its X-ray luminosity function and $P(\lambda | L_\mathrm{bol} ,z)$ is the distribution function of the Eddington ratio $\lambda$ in the form of log-normal distribution per unit $\mathrm{log}\lambda$ with fixed standard deviation $\sigma_{\log\lambda} = 0.3$ and averaged Eddington ratio of AGNs $\log \bar{\lambda} = -1.1 $ \citep{Ueda:2014tma}. The BHMF $\Phi_{\mathrm{BH}}$ can be obtained by solving the continuity equation of all SMBHs against the AGN mass function $\Psi_{\mathrm{AGN}}$ \citep{1992MNRAS.259..725S, Ueda:2014tma} given as
\begin{equation}
\frac{\partial \Phi_{\mathrm{BH}}(z,M)}{\partial z} \frac{dz}{dt} = -\frac{\partial}{\partial M} \left[ \frac{1-\eta(M)}{\eta(M)} \frac{\overline{\lambda}L_{\mathrm{Edd}}\Psi_\mathrm{AGN}(M,z)}{c^2} \right] \label{eq:continuity}
\end{equation}
where mass-dependent radiation efficiency in the form of $\eta(M) = 0.043\ (M/10^8 M_{\odot})^{0.54}$ is adopted \citep{Ueda:2014tma}. In this paper, we focus on SMBHs in the redshift range $0\leq z \leq 5$ and the mass range $10^5$-$10^{11} M_{\odot}$, where their demographics are well-constrained by the AGN X-ray luminosity functions \citep{Ueda:2014tma}. 

The merger timescale $\tau$ is evaluated as follows \citep{PhysRev.136.B1224,Quinlan:1996vp,2008gady.book.....B,Dosopoulou:2016hbg,DOrazio:2017dyb,Zhao:2023kff}. $\tau$ is modeled through four sequential stages: dynamical friction, stellar hardening, gaseous decay, and GW emission.

The first stage is driven by the dynamical friction given in the form of the resident timescale ($ t_{\rm res} = |a/\dot{a}|$ ) as \citep{Dosopoulou:2016hbg}

\begin{align}
t_{\rm DF} =\ & 0.12\ \mathrm{Gyr}
  \Bigg(\frac{a_{0}}{10\ \mathrm{kpc}}\Bigg)^2
  \Bigg(\frac{\sigma}{300\ \mathrm{km}\ \mathrm{s}^{-1}}\Bigg) \notag \\
 & \times \Bigg(\frac{q M}{10^8 M_{\odot}}\Bigg)^{-1}
  \frac{1}{\ln \Lambda}
\label{eq:dyn}
\end{align}
where $a_{0}$ is the initial orbital separation ,and $\sigma$ is the $M - \sigma$ scaling relation given in \citet{Tremaine:2002js}. We adopt the locally observed $M-\sigma$ relation throughout our analysis, noting that cumulative mass loss through repeated SMBH mergers could potentially modify this relation over cosmic time by up to $\sim 10-20\%$ \citep{Menou:2004by}. However, this effect is comparable to current observational uncertainties and would not qualitatively alter our conclusions. Here, ${\mathrm{ln} \Lambda}$ is the Coulomb logarithm defined as 
\begin{equation}
\Lambda = \frac{b_{\rm max}}{(G M_2)/ {v_{\rm DF}}^2}  
\end{equation}
where $M_2$ is the secondary BH mass, $b_{\rm max}$ is the maximum impact parameter, and $v_{\rm DF} =\sqrt{2} \sigma$ is the relative velocity of the secondary to the primary BH. We set an initial separation of 10~kpc as well as the impact parameter. While the dynamical friction timescale depends strongly on the initial separation, the binaries spend the majority of their time during the subsequent orbital decay. Therefore, though changing $a_0$ from 10 kpc to 40 kpc, corresponding to M87-like massive galaxy radius \citep{Cohen:1997tx} results in a factor of few difference in light of SGWB evaluation, the impact is limited in comparison with other parameters, such as binary gas accretion rate discussed below. For the sake of simplicity, we fix $a_0$ to its fiducial value in this study.  

When the orbital separation shrinks down to $\lesssim 10$ pc, the stellar hardening phase becomes dominant, formulated as \citep{Quinlan:1996vp,2008gady.book.....B}
\begin{equation}
t_{\rm h} = \frac{\sigma}{GH\rho_0 a} 
\label{eq:loss}
\end{equation}
where $H\ =14.3$ is the hardening rate coefficient \citep{2008gady.book.....B, Zhao:2023kff} and $\rho_0$ is the stellar density at the galaxy core. We adopt a fiducial value of $\rho_0 = 10.0 M_{\odot}/\rm{pc}^3$, representative of massive elliptical galaxies with stellar masses $M_{*}\sim10^{11}M_{\odot}$ that host the $10^{8-9} M_\odot$ SMBHs dominating the SGWB signal (see e.g., \citet{Sesana:2015haa, Goulding:2019hnn}).
                        
At the sub-pc scale, the depletion of stars renders stellar hardening inefficient, and the loss of orbital angular momentum through interactions with the surrounding circumbinary gas disk becomes the dominant mechanism driving the orbital decay of SMBHBs. While recent hydrodynamical simulations suggest that orbital evolution depends on binary and disk properties such as mass ratio, eccentricity, viscosity, and aspect ratio \citep[e.g.,][]{2020ApJ...900...43T, Valli:2024nbj, Duffell:2024fwy}, binaries in thin accretion disks expected for AGNs are found to undergo rapid inspiral driven by strong gravitational torques \citep{2020ApJ...900...43T}. The migration timescale derived from these thin-disk simulations is qualitatively consistent with analytical predictions based on angular momentum conservation. Given the uncertainties in constraining detailed disk parameters (e.g., viscosity, aspect ratio) for the SMBHB population, we adopt the analytical form as a fiducial description for gas-rich systems given as \citep{Loeb:2009rv, DOrazio:2017dyb}
\begin{equation}
t_{\rm gas} = \frac{q}{(1+q)^2}\frac{1}{\dot{\mathcal{M}}} t_{\rm Edd},\quad \dot{\mathcal{M}}=\frac{\dot{M}_{\rm tot}}{\dot{M}_{\rm Edd, tot}},
\label{eq:gas}
\end{equation}
where $\dot{\mathcal{M}}$ is the mass accretion rate of the total binary system ($\dot{M}_{\rm tot}$) normalized by the Eddington accretion rate ($\dot{M}_{\rm Edd, tot}$), and $t_{\rm Edd} \equiv M_{\rm tot}/\dot{M}_{\rm Edd,tot} \sim 4.5\times10^7$ yr is the Eddington timescale with a radiative efficiency of $0.1$. The migration timescale is inversely proportional to the normalized accretion rate because a higher gas supply enables more efficient extraction of orbital angular momentum from the binary, accelerating orbital decay \citep{Loeb:2009rv}. The mass ratio dependence arises because the total orbital angular momentum of the binary is proportional to the reduced mass, so binaries with more unequal masses have less angular momentum to lose and therefore migrate more slowly for a given accretion rate. We adopt $\dot{\mathcal{M}} = 0.1$ as our fiducial value, consistent with the average Eddington ratio of AGN found by \citet{Ueda:2014tma}. In the absence of direct measurements for dual AGN systems specifically, we assume their accretion properties are similar to those of the general AGN population, and we explore the impact of varying this assumption in Sec.~\ref{subsec:Gas_accretion}.

When the separation becomes small enough, the system starts to radiate GW emission. The coalescence timescale due to GW emission for a binary with eccentricity $e$ is given by \citep{PhysRev.136.B1224}
\begin{equation}
t_{\rm GW} = \frac{5c^5 a^4}{64G^3 M^3 q_s} \frac{1}{F(e)}
\label{eq:gw_eccentric}
\end{equation}
where $F(e)$ is the eccentricity factor denoted as \citep{PhysRev.136.B1224}
\begin{equation}
F(e) = \frac{1+(73/24)e^2+(37/96)e^4}{(1-e^2)^{7/2}}.
\end{equation}
We incorporate this eccentric orbital term to properly model the rapid decay in the final stage of the merger.

Each respective phase of SMBHB orbital evolution is defined by a characteristic effective radius, representing the orbital separation at which a specific physical process becomes the dominant mechanism driving the orbital shrinkage of the SMBHB (see \cite{Zhao:2023kff} for details). Consequently, the total merger timescale is evaluated as the sum of the solutions of Eqs. \ref{eq:dyn}, \ref{eq:loss}, \ref{eq:gas}, and \ref{eq:gw_eccentric}.

With the timescales defined above, we can explicitly determine the turnover frequency $f_t$. The frequency evolution rate driven by the gas disk interaction can be derived by assuming a Keplerian orbit as
\begin{equation}
\left( \frac{df}{dt} \right)_{\rm gas} = \frac{3}{2} \frac{f}{t_{\rm gas}}.
\end{equation}
On the other hand, the evolution rate due to GW emission is given by modifying the circular orbit formula with the enhancement factor $F(e)$ as \citep{Enoki:2006kj, Chen:2016zyo}
\begin{equation}
\left( \frac{df}{dt} \right)_{\rm GW} = \frac{96}{5} \pi^{8/3} \left( \frac{G \mathcal{M}}{c^3} \right)^{5/3} f^{11/3} F(e).
\end{equation}
The turnover frequency $f_t$ is derived from the equilibrium condition $(df/dt)_{\rm gas} = (df/dt)_{\rm GW}$. Solving this equality for $f$, we obtain
\begin{align} 
    f_t &= \left[ \frac{5}{64 \pi^{8/3} t_{\rm gas} F(e)} \left( \frac{c^3}{G \mathcal{M}} \right)^{5/3} \right]^{3/8} \notag \\ 
    &\simeq 2.2 \ \mathrm{nHz} \ \left( \frac{\dot{\mathcal{M}}}{0.1} \right)^{\frac{3}{8}} \left( \frac{\mathcal{M}}{10^9 M_\odot} \right)^{-\frac{5}{8}} \left( \frac{4\zeta}{1} \right)^{-\frac{3}{8}} \left( F(e) \right)^{-\frac{3}{8}}. 
\end{align}
By defining this turnover, we effectively incorporate the efficient orbital decay driven by gas inspirals for the modeling of the GW energy spectrum.

Our fiducial parameters for merger timescale is set as  $(a_0, e_0, \rho_0, \mathcal{\dot{M}})=(10\ \rm{kpc}, 0.45, 10.0\ M_{\odot}/ \rm{pc}^3, 0.1)$ to systematically assess our SGWB predictions. Of these parameters, the accretion rate $\mathcal{\dot{M}}$ exerts the strongest influence on the efficiency of SMBHB coalescence, and consequently, on the SGWB signal. We discuss the impact of $\mathcal{\dot{M}}$ in Sec.\ref{subsec:Gas_accretion}.
 
%%%%%%%%%%%%%%%%%%%%%%%%%%%%%%%%%%%%%%%%%%%
\section{Implications from dual AGN fraction}\label{sec:Dual_AGN}
\begin{figure}[t]
\includegraphics[width=1.0\linewidth]{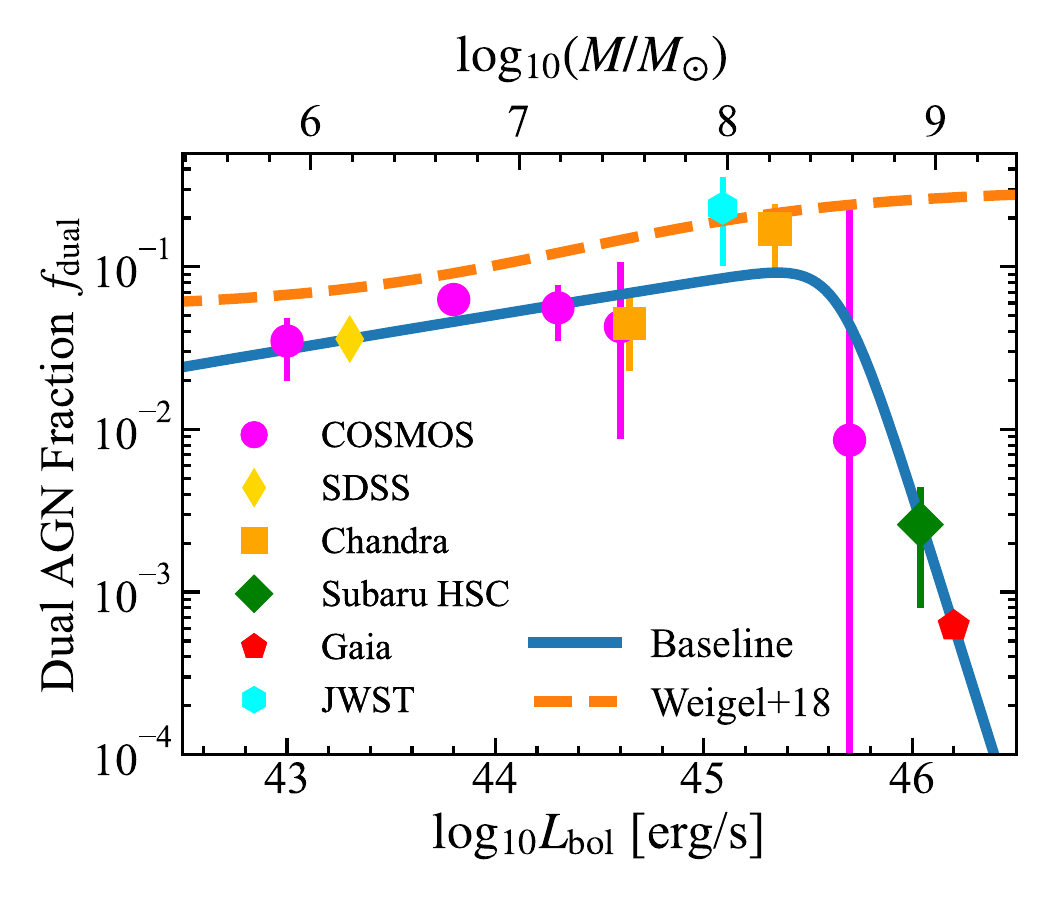} 
\caption{Dual AGN fraction as a function of bolometric luminosity and inferred black hole mass assuming $\lambda = 0.1$. Data points represent measurements from various surveys: COSMOS (circles) \citep{2024arXiv240514980L}, SDSS (thin diamonds) \citep{2011ApJ...737..101L}, Chandra (squares) \citep{2012ApJ...746L..22K}, Subaru HSC (diamonds) \citep{2020ApJ...899..154S}, Gaia (pentagons) \citep{Shen:2022cmp}, and JWST (hexagons) \citep{2023arXiv231003067P}. Lines show our baseline model (blue solid, fitted to data) and predictions from galaxy merger rates (orange dashed, "Weigel+18" \citep{2018MNRAS.476.2308W}). 
\label{fig:du_frac}}
\end{figure}

%%%%%%%%%%%%%%%%%%%%%%%%%%%%%%%%%%%%%%%%%%%
Recent observations have provided crucial insights into the prevalence of dual AGNs. Fig.~\ref{fig:du_frac} compiles observed dual AGN fraction $f_\mathrm{dual}$ from multiple surveys. Specifically, $f_{\rm dual}$ is derived from a parent sample with consistent luminosity, redshift, and spatial coverage selections. For instance, local surveys (SDSS at $0.02 < z < 0.16$ \citep{2011ApJ...737..101L}; Chandra X-ray at $z < 0.05$ \citep{2012ApJ...746L..22K}) probe large separations ($r_p < 100$ kpc) and moderate-luminosity AGNs ($L_{\rm bol} \sim 10^{43\text{--}45}~{\rm erg~s^{-1}}$), finding relatively high ($\sim 1-10 \%$) dual AGN fractions. In contrast, higher-redshift surveys focusing on luminous quasars ($L_{\rm bol} \gtrsim 10^{45}~{\rm erg~s^{-1}}$) with separations $r_p \lesssim 30$ kpc report lower fractions \citep{2020ApJ...899..154S, Shen:2022cmp}, while COSMOS ($z < 5$, $r_p < 15$ kpc) covering a broader luminosity range ($10^{43\text{--}46}~{\rm erg~s^{-1}}$) shows intermediate values \citep{2024arXiv240514980L}. Recent JWST observations suggest a notably higher fraction ($\sim$20\%) for $L_{\rm bol} \simeq 10^{44.5\text{--}46.0}~{\rm erg~s^{-1}}$ at $3 < r_p < 30$ kpc \citep{2023arXiv231003067P}. Regarding separation dependence, studies of nearby, moderate-luminosity AGNs find an increasing dual AGN fraction with decreasing separation \citep{2011ApJ...737..101L, 2012ApJ...746L..22K}. In contrast, surveys of high-luminosity quasars dominating the SGWB show relatively mild dependence on both redshift and separation within the observed range  \citep{2020ApJ...899..154S, Shen:2022cmp, 2024arXiv240514980L}. These weak trends with redshift and separation in the luminous quasar regime suggest a limited impact on our SGWB evaluation. The redshift dependence of $f_{\rm dual}$ is understood as a consequence of the stochastic, episodic nature of SMBH accretion during galaxy mergers \citep{Capelo:2016yrq, Goulding:2017mul}.  While the galaxy merger rate and gas content evolve over cosmic time \citep[e.g.,][]{Conselice:2014hqa, 2014AJ....148..137L}, the probability that both SMBHs in a merging system are simultaneously active as luminous AGN is primarily governed by short-term accretion variability rather than the global merger rate \citep{Capelo:2016yrq, Goulding:2017mul}. As a result, the dual AGN fraction would not be sensitive to cosmic evolution, though confirmation of this trend requires larger spectroscopic samples and a better understanding of selection effects, as discussed by \citet{2020ApJ...899..154S}.

Fig.~\ref{fig:du_frac} presents the luminosity dependence of the dual AGN fraction, which exhibits a clear decline at higher luminosities. This luminosity-dependent behavior can be attributed to several physical factors. Low-luminosity AGNs typically have longer duty cycles compared to more luminous AGNs \citep{2010MNRAS.406.1959S, 2013MNRAS.428..421S, Shen:2022cmp}. The longer activity duration naturally enhances the probability of detecting these systems as dual AGNs \citep{2011ApJ...737..101L, 2012ApJ...746L..22K}. Additionally, lighter binaries spend longer in the dynamical friction phase at kpc-scale separations before entering the sub-pc regime \citep{Begelman:1980vb}, providing an extended observational window for dual AGN detection in lower-luminosity systems.

We model this observed dual AGN fraction relation with a smoothly connected double power-law model to capture the observed characteristic luminosity dependence:
\begin{equation}
f_\mathrm{dual} (L_{\mathrm{bol}}) = A \left[\Big(\frac{L_{\mathrm{bol}}}{L_{*}}\Big)^{\gamma_1} + \Big(\frac{L_{\mathrm{bol}}}{L_{*}}\Big)^{\gamma_2} \right]^{-1} 
\end{equation}
where normalization constant $A$, break luminosity $L_{*}$ and slopes $\gamma_1, \gamma_2$ are fitted on the basis of $\chi^2$ algorithm. We obtain $\chi^2 = 5.7$ with a degree of freedom of 7, and the best-fit parameter values are $(A, \mathrm{log_{10}}L_{*}, \gamma_{1}, \gamma_{2}) = (0.113, 45.7, -0.211, 4.12)$. Here we also show dual AGN fraction as a function of $M$ in Fig. \ref{fig:du_frac} by assuming constant Eddington ratio $\lambda = 0.1$. For SGWB calculations, we convert this luminosity dependent $f_\mathrm{dual}$ to a mass-dependent fraction $f_\mathrm{dual} (M)$ by convolving with the Eddington ratio distribution  as:
\begin{equation}
f_\mathrm{dual} (M) = \int f_{\mathrm{dual}}(L_{\mathrm{bol}}) \frac{d \mathrm{log} L_{\mathrm{bol}}}{d \mathrm{log} M} P(\lambda | L_\mathrm{bol} ,z) d\mathrm{log}\lambda.
\end{equation}

For comparison, we also consider predictions from models that convert merging galaxy mass functions to AGN merger populations given in \citet{2018MNRAS.476.2308W} (dotted curve in Fig.~\ref{fig:du_frac}). Their model predicts increasing AGN merger fractions with luminosity due to the mass-dependent galaxy merger rate, where more massive galaxies are more likely to experience major mergers. The difference between the AGN pair fraction inferred from galaxy observations and our compiled dual AGN fraction could arise from the distinct physical processes each scenario traces. Here, the increasing trend of the galaxy major merger fraction with stellar mass is primarily driven by the non-linear stellar-to-halo mass relation (SHMR), as demonstrated by \citet{2017ApJ...845..145W}. In the low-mass regime, the SHMR is steep; consequently, even a major merger between dark matter halos often translates into a minor merger in stellar mass. In contrast, at the high-mass end where the SHMR flattens, halo mergers map more directly onto major stellar mergers, thereby enhancing the observed merger fraction for massive galaxies. While the galaxy-based pair fraction primarily reflects the cosmological evolution of galaxy mergers, where interactions can trigger AGN activity in either nucleus \citep{2012ApJ...758L..39T, 2018MNRAS.476.2308W}, the dual AGN fraction requires both SMBHs to be simultaneously active \citep{Capelo:2016yrq, Goulding:2017mul}.

Given these contrasting behaviors, we explore two scenarios for the dual AGN fraction in this paper: (i) our observationally motivated luminosity-dependent model (hereafter "Baseline") and (ii) the galaxy pair-based predictions of \cite{2018MNRAS.476.2308W} ("Weigel+18").

%%%%%%%%%%%%%%%%%%%%%%%%%%%%%%%%%%%%%%%%%%%
\section{Stochastic Gravitational Wave Background Spectrum}\label{sec:SGWB}
%%%%%%%%%%%%%%%%%%%%%%%%%%%%%%%%%%%%%%%%%%%
\begin{figure}[t]
\includegraphics[width=1.0\linewidth]{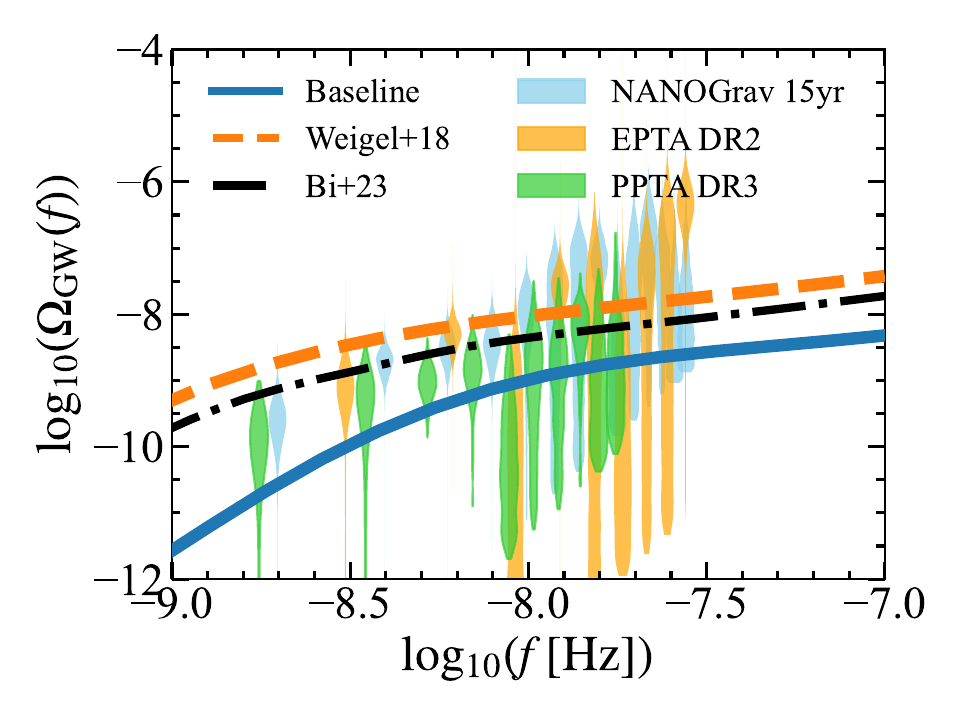} 
\caption{The SGWB energy density spectrum at PTA frequency bands. Model predictions using our baseline dual AGN fraction (blue solid) and galaxy pair-based fraction (orange dashed) are compared with the posterior distribution of the SGWB spectrum from NANOGrav 15yr (blue) \citep{NANOGrav:2023gor}, EPTA DR2 (orange) \citep{EPTA:2023fyk}, and PPTA DR3 (green) \citep{Reardon:2023gzh} measurements. The black dot-dashed curve shows a galaxy merger model fitted for the NANOGrav signals \citep{Bi:2023tib}.
\label{fig:SGWB}}
\end{figure}

We compute the present-day SGWB energy density for our two dual AGN fraction scenarios (Fig.~\ref{fig:SGWB}), compared with the posterior distribution of SGWB spectra from NANOGrav 15 years \citep{NANOGrav:2023gor}, European PTA Data Release 2 (EPTA DR2) \citep{EPTA:2023fyk}, and Parkes PTA Data Release 3 (PPTA DR3) \citep{Reardon:2023gzh}. For comparison, we show results from a phenomenological galaxy evolution model derived via MCMC sampling of the NANOGrav data \citep{Bi:2023tib}. It is important to emphasize that while our analytical model predicts a smooth spectrum representing the ensemble average, the actual SGWB at these high frequencies ($f \gtrsim 10^{-8}$~Hz) is dominated by a finite number of discrete sources. This discreteness leads to significant realization variance, causing the actual spectrum to appear jagged rather than smooth \citep[e.g.,][]{Sesana:2008xk, Rosado:2015epa, Kelley:2017vox}. Although modeling these specific realizations via Monte Carlo sampling is beyond the scope of this work, our result provides the expected amplitude of the background assuming a continuous population.

Our Baseline model case shows good agreement with the observed nHz SGWB signals across the PTA frequency range. In contrast, the galaxy pair-based model ("Weigel+18") overproduces the low-frequency signals at $\lesssim 3\mathrm{nHz}$ due to abundant $\gtrsim 10^9 M_{\odot}$ SMBHB population. These results demonstrate that our observationally motivated baseline model better captures the spectral behavior of the SGWB.

%%%%%%%%%%%%%%%%%%%%%%%%%%%%%%%%%%%%%%%%%%%
\section{Discussions}\label{sec:Discussion}
%%%%%%%%%%%%%%%%%%%%%%%%%%%%%%%%%%%%%%%%%%%
\subsection{Impact of Gas Accretion onto the Binaries}\label{subsec:Gas_accretion}
\begin{figure}[t]
\includegraphics[width=1.0\linewidth]{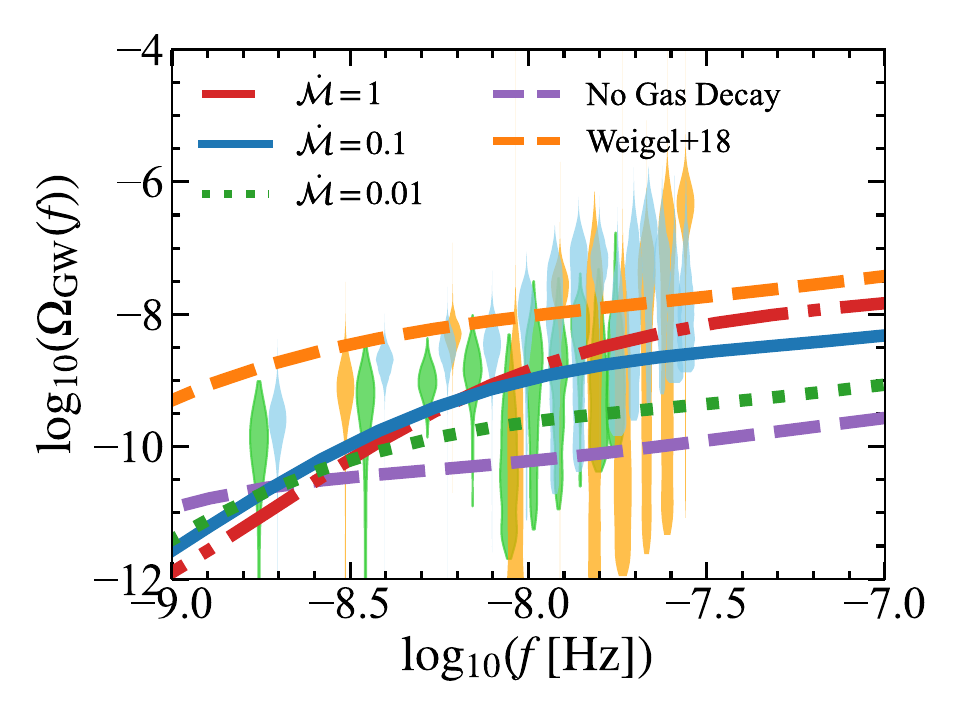} 
\caption{The SGWB energy density spectrum at PTA frequency regimes, illustrating the impact of varying the accretion rate during the gaseous decay phase. The dash-dotted red, solid blue, and dotted green curves represent models with Eddington accretion rates $\mathcal{\dot{M}} = 1, 0.1$ (Baseline), and $0.01$ respectively. The dashed purple curve shows the scenario without gas decay. The orange dashed curve corresponds to the galaxy pair model. The posterior distribution of the observed SGWB spectrum is also shown in the same format as Fig. \ref{fig:SGWB}. All parameters other than $\mathcal{\dot{M}}$ are assumed to be fiducial.
\label{fig:Gas_decay}}
\end{figure}

The influence of the accretion rate during the gaseous decay phase on the predicted SGWB spectrum is illustrated in Fig. \ref{fig:Gas_decay}. Here, the spectral turnover is determined by the transition from the gas-driven phase to the GW-driven phase. A higher accretion rate implies more efficient angular momentum transport, causing spectrum turnover at higher frequencies. Consequently, increasing the accretion rate results in a more pronounced suppression of the SGWB amplitude in the nHz regime. As shown in Fig. \ref{fig:Gas_decay}, the model with our fiducial rate $\dot{\mathcal{M}}=0.1$ exhibits a turnover around several nHz. Reducing the accretion rate to $\dot{\mathcal{M}}=0.01$ lowers the overall amplitude by a factor of several due to the inefficient orbital decay, potentially leading to a fraction of binaries stalling before entering the GW-dominated regime. Conversely, a high accretion rate of $\dot{\mathcal{M}}=1.0$ produces the highest amplitude in the high-frequency regime but drops off steeply at $f \lesssim 10$ nHz due to the rapid gas-driven inspiral evoltion. This highlights the critical role of gaseous decay in shaping the SGWB spectrum, particularly in determining the turnover frequency and modifying the power-law behavior in the nHz band \citep[see e.g.,][and references therein]{Lai:2022ylu, Duffell:2024fwy}.

The existence of gas-driven orbital decay fundamentally relies on the assumption that SMBHBs are embedded in a gas-rich environment and experience active accretion during a certain portion of their merger timescale. This assumption is supported since galaxy mergers provide
plenty of gas toward the center of the merger remnant \citep{1996ApJ...471..115B, Hopkins:2005fb}, leading to the formation of a gaseous circumbinary disk around SMBHBs. To robustly constrain the migration timescales and accretion processes during the gaseous decay phase, coordinated efforts combining high-resolution multiwavelength observations, advanced hydrodynamic simulations, and theoretical modeling are essential for achieving a comprehensive understanding of SMBHB coalescence and its contribution to the nHz gravitational wave background \citep{Lai:2022ylu}.

\subsection{Eccentric Evolution of Inspiraling Binaries}\label{subsec:eccentricity}
\begin{figure}[t]
\includegraphics[width=1.0\linewidth]{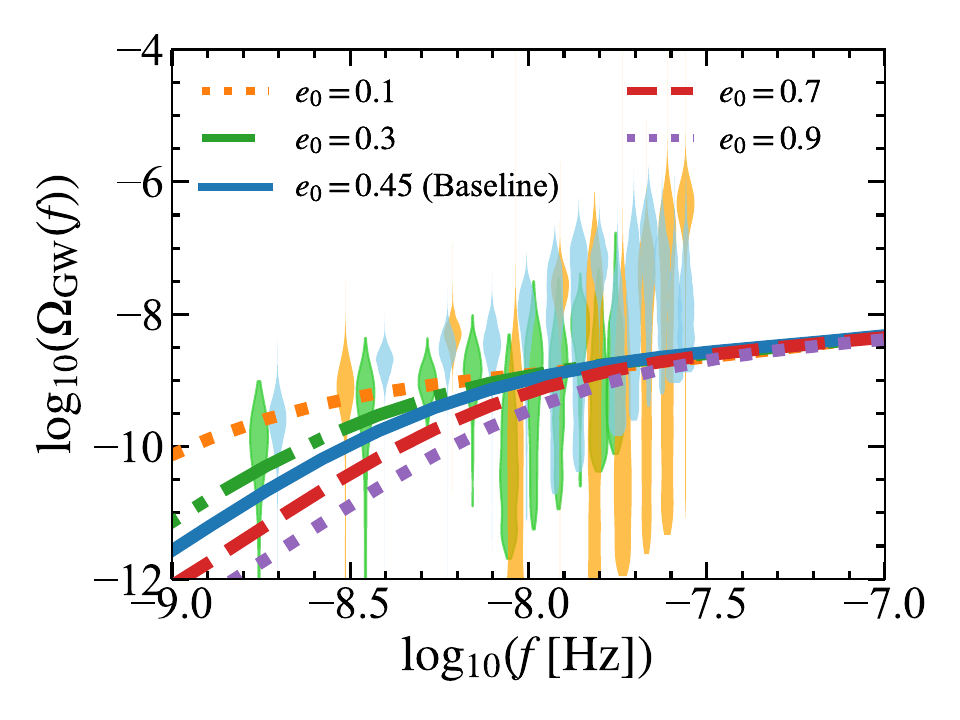} 
\caption{The energy density spectrum of the SGWB in the nHz GW regime, illustrating the impact of varying the initial eccentricity at the inspiral phase. The dotted (orange), dashed (green), solid (blue), dash-dotted (red), and dotted (purple) curves represent models with initial eccentricities $e_0=0.1,0.3, 0.45\ (\rm Baseline), 0.7$, and $0.9$, respectively. The posterior distribution of the observed SGWB spectrum is also shown in the same way as in Fig. \ref{fig:SGWB}. All parameters other than $e_0$ are assumed to be fiducial as $(a_0, \mathcal{\dot{M}}, \rho_0)=(10\ \rm{kpc}, 0.1, 10.0\ M_{\odot}/ \rm{pc}^3)$.   
\label{fig:ecc}}
\end{figure}

The eccentricity of SMBHBs has been recognized as an important factor that can modify the SGWB spectrum in the low frequency regime \citep{Sesana:2013wja, Ravi:2014aha, Kelley:2017lek, Moreschi:2025qtm, Sah:2025dmv}. We further investigated the impact of the initial eccentricity at the beginning of the GW emission phase in Fig. \ref{fig:ecc}. Since the spectral turnover is governed by the decoupling from the gas disk, which occurs at relatively high frequencies ($f \sim$ nHz) in our gas-driven model, the impact of eccentricity extends into the frequency band relevant for PTA observations. As seen in Fig. \ref{fig:ecc}, increasing the initial eccentricity significantly suppresses the SGWB amplitude not only at sub-nHz frequencies but also around the turnover frequency in the nHz regime. For instance, at $f \sim 3$ nHz, the model with high eccentricity ($e_0 = 0.9$) shows a suppression of approximately 2 dex compared to the low eccentricity case ($e_0 = 0.1$). This suppression arises from two effects: (1) highly eccentric binaries emit GWs more efficiently, evolving faster and thus spending less time in each frequency bin, which reduces the number of sources contributing to the background \citep{Kelley:2017lek}, and (2) eccentricity redistributes GW power from the fundamental orbital frequency to higher harmonics, flattening the spectrum at lower frequencies \citep{Sesana:2013wja}.

Nevertheless, the turnover feature at low frequencies will be crucial for future observations, as it can serve as a diagnostic for the eccentricity distribution in the SMBHB population and the physical processes governing binary evolution. The circumbinary gas disks are thought to play a key role in determining eccentricity. Current theoretical studies suggest that prograde disks can either circularize ($e_0\sim0$), or, under certain conditions, settle to an equilibrium state around $e_0 \sim 0.4 $ \citep{Zrake:2020zkw, DOrazio:2021kob, Siwek:2023rlk}. In contrast, retrograde disks are predicted to pump eccentricities to even higher values up to $e_0 \geq 0.8$ \citep{Garg:2024oeu, 2024MNRAS.527.6021T}. Consequently, the disk properties directly dictate the degree of eccentricity, determining which of the suppression mechanisms described above dominates the spectral shape. Thus, the resultant overall signature of these SMBHB evolutions within the gaseous decay phase could be imprinted on the SGWB spectrum.

However, environmental effects from gas and stellar interactions can also suppress the low-frequency SGWB \citep{Ellis:2023dgf, Raidal:2024odr}, producing spectral features that can be difficult to distinguish from eccentricity-induced suppression. Distinguishing eccentricity from these environmental factors will require future detailed analysis using techniques such as frequency-binned correlations \citep{Raidal:2024odr}.

\subsection{Observational Constraints and Selection Effects on SMBHB Demographics}\label{sebsec:observed}
The observational detection of dual AGNs is crucial for bridging EM and GW observations and for testing models of SMBHB evolution. In this paper, we consider that SMBHBs are expected to undergo multiple active phases throughout their merger timescale, which typically spans a few hundred million to a few billion years. Since individual quasar lifetimes are estimated to be much shorter ($10^6$-$10^8$ years) \citep{Martini:2003ek}, the observed fraction of dual AGNs can be considered an average snapshot of these intermittent phases. This allows us to use the observed dual AGN fraction as a viable proxy for the underlying SMBHB population, including inactive binaries. A key assumption underlying our approach is that the observed dual AGN fraction $f_\mathrm{dual}$ traces the subset of SMBH binaries most relevant to the nHz SGWB. This requires coupled activation rather than independent SMBH activity; if triggering were purely stochastic, $f_{\rm dual}$ would be suppressed by the square of the duty cycle, making simultaneous activity extremely rare. Recent theoretical analyses combined with observational constraints suggest this triggering behavior. \citet{Padmanabhan:2024nvv} found that consistency between JWST dual AGN observations and NANOGrav's SGWB limits requires both BHs at kpc scales to be powered by shared gas reservoirs rather than activating independently. \citet{Casey-Clyde:2024hwg} performed a multi-messenger analysis combining NANOGrav's measurements with the periodic quasar catalog, demonstrating that quasars are up to 5 times more likely to host SMBHBs than random galaxies, indicating SMBHBs are more likely to be activated. Furthermore, \citet{2010MNRAS.407.1529H} showed through hydrodynamical simulations that the same large-scale gravitational torques that channel gas from kpc to sub-pc scales can trigger coherent activation of both SMBHs. Our result provides additional context: our dual AGN-based model reproduces the observed SGWB, interpreting $f_\mathrm{dual}$ as tracing the subset of binaries undergoing efficient gas-driven inspiral, implying dual AGN are likely to be the dominant SGWB sources over non-activated ones. We acknowledge that the precise relationship between $f_{\rm dual}$ and the underlying binary population remains to be established through expanded SMBH pair observations and improved theoretical modeling, including large-scale cosmological simulations. 

As for the observational selections of dual AGN samples, current observational methods are subject to certain selection biases that limit our understanding of the underlying SMBHB population. For instance, the wavelength-dependent selection effects bias samples, with mid-infrared methods preferentially selecting dusty, star-forming AGN and X-ray methods favoring gas-poor systems \citep{2017ApJ...835...27A, 2020ApJ...888...78S}. Additionally, most dual AGN searches are sensitive only to projected separations of $\sim 10$ kpc due to current angular resolution limits. However, dual AGNs span a broad separation distribution at each redshift, and observations limited to this range will systematically miss closer pairs. This separation-dependent selection bias makes it difficult to predict what fraction of observed dual AGNs will successfully evolve into gravitationally-bound SMBHBs and suggests that modeling of the separation distribution is inherently required to robustly link observed $f_\mathrm{dual}$ to the underlying SMBHB population. 

Next-generation X-ray missions (Athena \citep{2013arXiv1306.2307N}, Lynx \citep{2018arXiv180909642T}, AXIS \citep{Foord:2023fdj}) will improve sensitivity to obscured dual AGNs at ~kpc to ~100 pc separations, while sub-mm very long baseline interferometry (VLBI) will directly image the inner sub-pc region \citep{DOrazio:2017dyb, Zhao:2023kff}. Next‐generation optical and near‐infrared facilities—including wide‐field surveys such as the Vera C. Rubin Observatory \citep{LSST:2008ijt} and JWST, together with extremely large telescopes (e.g., Thirty Meter Telescope \citep{TMTInternationalScienceDevelopmentTeamsTMTScienceAdvisoryCommittee:2015pvw}, Giant Magellan Telescope \citep{2012SPIE.8444E..1HJ}, European Extremely Large Telescope \citep{Padovani:2023dxc})—will expand dual AGN samples, enabling both broad statistical studies and high‐resolution follow‐up of close pairs. In parallel, PTAs and future radio arrays (Square Kilometre Array (SKA) and Next-Generation Very Large Array (ngVLA)) will resolve individual SMBH binaries in the nHz band \citep{2017NatAs...1..886M, Kelley:2017vox, Becsy:2022pnr, Garg:2024oeu}. Recent NANOGrav analyses have already identified two marginal SMBHB candidates in the NANOGrav 15-year dataset \citep{Agarwal:2025cag}. This multi-messenger synergy will span sub-pc to kpc scales, linking EM dual AGN signatures to GW detections and tightly constraining SMBHB demographics.

\subsection{Comparison with Previous Works}\label{sebsec:comparison}
The SGWB, predominantly generated by SMBH mergers, has been the subject of numerous theoretical predictions in the literature. These predictions for the SGWB originate from several distinct methodologies. The standard methodology relies on either SAMs or cosmological simulations of galaxy evolution. Models based on galaxies, for instance, utilize a comprehensive framework to estimate the galaxy merger rate from the galaxy stellar mass function and pair fraction, in conjunction with the SMBH-galaxy mass scaling relation \citep{Jaffe:2002rt, Sesana:2012ak, Sesana:2016yky, Chen:2020qlp}. In cosmological simulations, the SGWB spectrum is computed based on the evolution and population distribution of SMBHBs, where the unresolved region is modeled using post-processing for SMBHB evolution \citep{Ravi:2014aha, Kelley:2016gse, Saeedzadeh:2024wkz}. 

More recently, alternative approaches have emerged. Quasar-based models infer the SMBHB population from quasar properties, such as their luminosity functions or quasar lifetime, detouring the modeling of the galaxy mergers from simulations or SAMs \citep{2009ApJ...700.1952H, Goulding:2019hnn, Xin:2021mmk, Casey-Clyde:2021xro, Kis-Toth:2024gkm, Xin:2025voy, Lapi:2025wxt}. For instance, \citet{Goulding:2019hnn} used the discovery of a specific dual AGN as an empirical anchor for SGWB predictions, suggesting that such systems could contribute 1-10 \% of the SGWB signal. \citet{Casey-Clyde:2021xro} created a model that assumes the number of quasars and SMBHBs is proportional, using the measured SGWB amplitude to predict the local number of SMBHBs. Similarly, \citet{Kis-Toth:2024gkm} calculates the SGWB from the quasar luminosity function by assuming that quasar activity is triggered by mergers, finding that a scenario where all quasars are associated with SMBHBs yields a SGWB amplitude consistent with the NANOGrav.

Our approach differs from previous models in its physical assumptions. For instance, models based on galaxies often rely on indirect theoretical and empirical assumptions to infer the SMBHB population, such as the galaxy pair formation rate, which can introduce uncertainties between the galaxy pairs and the actual SMBHB population. Similarly, cosmological simulations, despite providing a comprehensive view of galaxy formation and SMBH mergers, the predicted SGWB often depends heavily on the simulation settings, including the specific sub-grid choices and the simulation's resolution. Our model, in contrast, offers a more observationally grounded approach by directly incorporating the constrained AGN observables.

Meanwhile, the quasar-based models rely on less constrained assumptions about the relationship between SMBHBs and quasars. For instance, \cite{Kis-Toth:2024gkm} assumes a simplified one-to-one correspondence where all quasars are associated with SMBHBs. Our methodology, by leveraging luminosity-dependent dual AGN demographics, provides a more specific and empirically grounded connection. This point should also be compared with \cite{Padmanabhan:2024nvv}. Their model computes the SGWB based on the merger rate of dark matter halos combined with the observed dual AGN fraction confirmed by recent JWST observations, leading to the conclusion that a high fraction ($\sim 20-30 \%$) alone can lead to overshooting the observed NANOGrav spectrum. Our approach, by contrast, utilizes a broader luminosity range of observational data, which allows for reconciliation with the observed GW data. 

Some recent empirical models have highlighted the challenge of explaining the PTA-measured SGWB amplitude. \citet{Sato-Polito:2023gym} found that predictions based on local SMBH demographics underestimate the observed signal, requiring an undetected population of massive ($>10^9 M_{\odot}$) BHs. Likewise, \citet{Lapi:2025wxt} developed a semi-empirical framework that self-consistently models SMBH accretion and mergers, concluding that SMBHB contributions can account for 30–50\% of observed SGWB amplitude at most. We note that our approach differs in that we also track the population-level evolution of SMBHBs, including merger-driven depletion, the evolution of the pair fraction, and merger-timescale scalings, i.e., the overall population budget. These ingredients can shift the predicted SGWB amplitude by a factor of $\lesssim 2$ relative to the fiducial value in \citet{Sato-Polito:2023gym}. The remaining tension should be viewed as mild given its sensitivity to pulsar-noise modeling; for instance, \citet{Goncharov:2024htb} demonstrated that improved noise analysis yields a lower intrinsic SGWB amplitude, bringing the observational constraints closer to the predictions of empirical SMBH mass functions.

%%%%%%%%%%%%%%%%%%%%%%%%%%%%%%%%%%%%%%%%%%%
\section{Conclusions}\label{sec:Conclusion}
%%%%%%%%%%%%%%%%%%%%%%%%%%%%%%%%%%%%%%%%%%%
Our analysis, combining dual AGN statistics and pulsar timing array gravitational wave signals, provides new insights into SMBHB populations. We find a distinct luminosity dependence in dual AGN fractions that decreases at higher luminosities in the massive ($\geq 10^9 M_{\odot}$) regime, contrasting with predictions from major galaxy merger pair fractions. The observed nHz SGWB signals show consistency with SMBHB demographics when incorporating our luminosity-dependent dual AGN fractions combined with AGN evolution, while the galaxy pair model would overproduce the signal.

The expected SGWB spectrum could depend on the physical mechanisms of orbital evolution. Especially, the gas accretion rate and the eccentricity of the binaries play pivotal roles in shaping the SGWB spectrum. Our model demonstrates that gas-driven orbital decay is essential for explaining the observed SGWB signal, as a scenario without this process is insufficient to reproduce current observational data. The accretion rate of the binaries primarily impacts the normalization of the SGWB spectrum, which is the overall amplitude, with higher accretion rates leading to larger SGWB signals. In contrast, the initial eccentricity of the binaries has a notable impact on the shape of the SGWB spectrum, particularly at very low frequencies, influencing the spectral turnover feature at nHz frequencies and below. This distinction highlights how different physical processes leave unique imprints on the SGWB signal, providing diagnostic tools for understanding the properties of SMBHB populations. These physical clues implicated by this work can be tested by both future electromagnetic and GW observations. \\

\section*{Acknowledgements}
We thank the anonymous referee for constructive comments that significantly improved the manuscript. We also would like to thank Keitaro Takahashi for useful discussions and comments. YI is supported by NAOJ ALMA Scientific Research Grant Number 2021-17A; JSPS KAKENHI Grant Number JP18H05458, JP19K14772, and JP22K18277; and World Premier International Research Center Initiative (WPI), MEXT, Japan. DT was supported in part by the JSPS Grant-in-Aid for Scientific Research (22K21349).

\bibliography{references}{}

@article{NANOGrav:2023hde,
    author = "Agazie, Gabriella and others",
    collaboration = "NANOGrav",
    title = "{The NANOGrav 15 yr Data Set: Observations and Timing of 68 Millisecond Pulsars}",
    eprint = "2306.16217",
    archivePrefix = "arXiv",
    primaryClass = "astro-ph.HE",
    doi = "10.3847/2041-8213/acda9a",
    journal = "Astrophys. J. Lett.",
    volume = "951",
    number = "1",
    pages = "L9",
    year = "2023"
}

@article{NANOGrav:2023gor,
    author = "Agazie, Gabriella and others",
    collaboration = "NANOGrav",
    title = "{The NANOGrav 15 yr Data Set: Evidence for a Gravitational-wave Background}",
    eprint = "2306.16213",
    archivePrefix = "arXiv",
    primaryClass = "astro-ph.HE",
    doi = "10.3847/2041-8213/acdac6",
    journal = "Astrophys. J. Lett.",
    volume = "951",
    number = "1",
    pages = "L8",
    year = "2023"
}

@article{EPTA:2023fyk,
    author = "Antoniadis, J. and others",
    collaboration = "EPTA, InPTA:",
    title = "{The second data release from the European Pulsar Timing Array - III. Search for gravitational wave signals}",
    eprint = "2306.16214",
    archivePrefix = "arXiv",
    primaryClass = "astro-ph.HE",
    doi = "10.1051/0004-6361/202346844",
    journal = "Astron. Astrophys.",
    volume = "678",
    pages = "A50",
    year = "2023"
}

@article{EPTA:2023xxk,
    author = "Antoniadis, J. and others",
    collaboration = "EPTA, InPTA",
    title = "{The second data release from the European Pulsar Timing Array - IV. Implications for massive black holes, dark matter, and the early Universe}",
    eprint = "2306.16227",
    archivePrefix = "arXiv",
    primaryClass = "astro-ph.CO",
    doi = "10.1051/0004-6361/202347433",
    journal = "Astron. Astrophys.",
    volume = "685",
    pages = "A94",
    year = "2024"
}

@article{Reardon:2023zen,
    author = "Reardon, Daniel J. and others",
    title = "{The Gravitational-wave Background Null Hypothesis: Characterizing Noise in Millisecond Pulsar Arrival Times with the Parkes Pulsar Timing Array}",
    eprint = "2306.16229",
    archivePrefix = "arXiv",
    primaryClass = "astro-ph.HE",
    doi = "10.3847/2041-8213/acdd03",
    journal = "Astrophys. J. Lett.",
    volume = "951",
    number = "1",
    pages = "L7",
    year = "2023"
}

@article{Reardon:2023gzh,
    author = "Reardon, Daniel J. and others",
    title = "{Search for an Isotropic Gravitational-wave Background with the Parkes Pulsar Timing Array}",
    eprint = "2306.16215",
    archivePrefix = "arXiv",
    primaryClass = "astro-ph.HE",
    doi = "10.3847/2041-8213/acdd02",
    journal = "Astrophys. J. Lett.",
    volume = "951",
    number = "1",
    pages = "L6",
    year = "2023"
}

@article{Xu:2023wog,
    author = "Xu, Heng and others",
    title = "{Searching for the Nano-Hertz Stochastic Gravitational Wave Background with the Chinese Pulsar Timing Array Data Release I}",
    eprint = "2306.16216",
    archivePrefix = "arXiv",
    primaryClass = "astro-ph.HE",
    doi = "10.1088/1674-4527/acdfa5",
    journal = "Res. Astron. Astrophys.",
    volume = "23",
    number = "7",
    pages = "075024",
    year = "2023"
}

@article{Vilenkin:1984ib,
    author = "Vilenkin, Alexander",
    title = "{Cosmic Strings and Domain Walls}",
    reportNumber = "PRINT-84-0840 (TUFTS)",
    doi = "10.1016/0370-1573(85)90033-X",
    journal = "Phys. Rept.",
    volume = "121",
    pages = "263--315",
    year = "1985"
}

@article{Hindmarsh:1994re,
    author = "Hindmarsh, M. B. and Kibble, T. W. B.",
    title = "{Cosmic strings}",
    eprint = "hep-ph/9411342",
    archivePrefix = "arXiv",
    reportNumber = "SUSX-TP-94-74, IMPERIAL-TP-94-95-5, NI-94025",
    doi = "10.1088/0034-4885/58/5/001",
    journal = "Rept. Prog. Phys.",
    volume = "58",
    pages = "477--562",
    year = "1995"
}

@article{Caprini:2015zlo,
    author = "Caprini, Chiara and others",
    title = "{Science with the space-based interferometer eLISA. II: Gravitational waves from cosmological phase transitions}",
    eprint = "1512.06239",
    archivePrefix = "arXiv",
    primaryClass = "astro-ph.CO",
    reportNumber = "DESY-15-246",
    doi = "10.1088/1475-7516/2016/04/001",
    journal = "JCAP",
    volume = "04",
    pages = "001",
    year = "2016"
}

@article{Guzzetti:2016mkm,
    author = "Guzzetti, M. C. and Bartolo, N. and Liguori, M. and Matarrese, S.",
    title = "{Gravitational waves from inflation}",
    eprint = "1605.01615",
    archivePrefix = "arXiv",
    primaryClass = "astro-ph.CO",
    doi = "10.1393/ncr/i2016-10127-1",
    journal = "Riv. Nuovo Cim.",
    volume = "39",
    number = "9",
    pages = "399--495",
    year = "2016"
}

@article{Saikawa:2017hiv,
    author = "Saikawa, Ken'ichi",
    title = "{A review of gravitational waves from cosmic domain walls}",
    eprint = "1703.02576",
    archivePrefix = "arXiv",
    primaryClass = "hep-ph",
    reportNumber = "DESY-17-036",
    doi = "10.3390/universe3020040",
    journal = "Universe",
    volume = "3",
    number = "2",
    pages = "40",
    year = "2017"
}

@article{Caprini:2019egz,
    author = "Caprini, Chiara and others",
    title = "{Detecting gravitational waves from cosmological phase transitions with LISA: an update}",
    eprint = "1910.13125",
    archivePrefix = "arXiv",
    primaryClass = "astro-ph.CO",
    reportNumber = "DESY-19-159, IPPP/19/27, HIP-2019-14/TH, MITP/19-066, IFT-UAM/CSIC-19-139",
    doi = "10.1088/1475-7516/2020/03/024",
    journal = "JCAP",
    volume = "03",
    pages = "024",
    year = "2020"
}

@article{Hindmarsh:2020hop,
    author = {Hindmarsh, Mark B. and L\"uben, Marvin and Lumma, Johannes and Pauly, Martin},
    title = "{Phase transitions in the early universe}",
    eprint = "2008.09136",
    archivePrefix = "arXiv",
    primaryClass = "astro-ph.CO",
    reportNumber = "MPP-2020-163, HIP-2020-27/TH",
    doi = "10.21468/SciPostPhysLectNotes.24",
    journal = "SciPost Phys. Lect. Notes",
    volume = "24",
    pages = "1",
    year = "2021"
}

@article{Domenech:2021ztg,
    author = "Dom\`enech, Guillem",
    title = "{Scalar Induced Gravitational Waves Review}",
    eprint = "2109.01398",
    archivePrefix = "arXiv",
    primaryClass = "gr-qc",
    doi = "10.3390/universe7110398",
    journal = "Universe",
    volume = "7",
    number = "11",
    pages = "398",
    year = "2021"
}

@article{Yuan:2021qgz,
    author = "Yuan, Chen and Huang, Qing-Guo",
    title = "{A topic review on probing primordial black hole dark matter with scalar induced gravitational waves}",
    eprint = "2103.04739",
    archivePrefix = "arXiv",
    primaryClass = "astro-ph.GA",
    doi = "10.1016/j.isci.2021.102860",
    journal = "iScience",
    volume = "24",
    pages = "102860",
    year = "2021"
}

@article{NANOGrav:2023hvm,
    author = "Afzal, Adeela and others",
    collaboration = "NANOGrav",
    title = "{The NANOGrav 15 yr Data Set: Search for Signals from New Physics}",
    eprint = "2306.16219",
    archivePrefix = "arXiv",
    primaryClass = "astro-ph.HE",
    reportNumber = "FERMILAB-PUB-23-589-T",
    doi = "10.3847/2041-8213/acdc91",
    journal = "Astrophys. J. Lett.",
    volume = "951",
    number = "1",
    pages = "L11",
    year = "2023",
    note = "[Erratum: Astrophys.J.Lett. 971, L27 (2024), Erratum: Astrophys.J. 971, L27 (2024)]"
}

@article{Rajagopal:1994zj,
    author = "Rajagopal, Mohan and Romani, Roger W.",
    title = "{Ultralow frequency gravitational radiation from massive black hole binaries}",
    eprint = "astro-ph/9412038",
    archivePrefix = "arXiv",
    doi = "10.1086/175813",
    journal = "Astrophys. J.",
    volume = "446",
    pages = "543--549",
    year = "1995"
}

@article{Jaffe:2002rt,
    author = "Jaffe, Andrew H. and Backer, Donald C.",
    title = "{Gravitational waves probe the coalescence rate of massive black hole binaries}",
    eprint = "astro-ph/0210148",
    archivePrefix = "arXiv",
    doi = "10.1086/345443",
    journal = "Astrophys. J.",
    volume = "583",
    pages = "616--631",
    year = "2003"
}

@article{Wyithe:2002ep,
    author = "Wyithe, J. Stuart B. and Loeb, Abraham",
    title = "{Low - frequency gravitational waves from massive black hole binaries: Predictions for LISA and pulsar timing arrays}",
    eprint = "astro-ph/0211556",
    archivePrefix = "arXiv",
    doi = "10.1086/375187",
    journal = "Astrophys. J.",
    volume = "590",
    pages = "691--706",
    year = "2003"
}

@article{Sesana:2004sp,
    author = "Sesana, Alberto and Haardt, Francesco and Madau, Piero and Volonteri, Marta",
    title = "{Low - frequency gravitational radiation from coalescing massive black hole binaries in hierarchical cosmologies}",
    eprint = "astro-ph/0401543",
    archivePrefix = "arXiv",
    doi = "10.1086/422185",
    journal = "Astrophys. J.",
    volume = "611",
    pages = "623--632",
    year = "2004"
}

@article{Enoki:2004ew,
    author = "Enoki, Motohiro and Inoue, Kaiki Taro and Nagashima, Masahiro and Sugiyama, Naoshi",
    title = "{Gravitational waves from supermassive black hole coalescence in a hierarchical galaxy formation model}",
    eprint = "astro-ph/0404389",
    archivePrefix = "arXiv",
    doi = "10.1086/424475",
    journal = "Astrophys. J.",
    volume = "615",
    pages = "19",
    year = "2004"
}

@article{Enoki:2006kj,
    author = "Enoki, Motohiro and Nagashima, Masahiro",
    title = "{The Effect of Orbital Eccentricity on Gravitational Wave Background Radiation from Cosmological Binaries}",
    eprint = "astro-ph/0609377",
    archivePrefix = "arXiv",
    doi = "10.1143/PTP.117.241",
    journal = "Prog. Theor. Phys.",
    volume = "117",
    pages = "241",
    year = "2007"
}

@article{Burke-Spolaor:2018bvk,
    author = "Burke-Spolaor, Sarah and others",
    title = "{The Astrophysics of Nanohertz Gravitational Waves}",
    eprint = "1811.08826",
    archivePrefix = "arXiv",
    primaryClass = "astro-ph.HE",
    doi = "10.1007/s00159-019-0115-7",
    journal = "Astron. Astrophys. Rev.",
    volume = "27",
    number = "1",
    pages = "5",
    year = "2019"
}

@article{Sesana:2008xk,
    author = "Sesana, A. and Vecchio, A. and Volonteri, M.",
    title = "{Gravitational waves from resolvable massive black hole binary systems and observations with Pulsar Timing Arrays}",
    eprint = "0809.3412",
    archivePrefix = "arXiv",
    primaryClass = "astro-ph",
    doi = "10.1111/j.1365-2966.2009.14499.x",
    journal = "Mon. Not. Roy. Astron. Soc.",
    volume = "394",
    pages = "2255",
    year = "2009"
}

@article{Sesana:2013wja,
    author = "Sesana, A.",
    title = "{Insights into the astrophysics of supermassive black hole binaries from pulsar timing observations}",
    eprint = "1307.2600",
    archivePrefix = "arXiv",
    primaryClass = "astro-ph.CO",
    doi = "10.1088/0264-9381/30/22/224014",
    journal = "Class. Quant. Grav.",
    volume = "30",
    pages = "224014",
    year = "2013"
}

@article{Ravi:2014aha,
    author = "Ravi, V. and Wyithe, J. S. B. and Shannon, R. M. and Hobbs, G. and Manchester, R. N.",
    title = "{Binary supermassive black hole environments diminish the gravitational wave signal in the pulsar timing band}",
    eprint = "1404.5183",
    archivePrefix = "arXiv",
    primaryClass = "astro-ph.CO",
    doi = "10.1093/mnras/stu779",
    journal = "Mon. Not. Roy. Astron. Soc.",
    volume = "442",
    number = "1",
    pages = "56--68",
    year = "2014"
}

@ARTICLE{2011MNRAS.411.1467K,
       author = {{Kocsis}, B. and {Sesana}, A.},
        title = "{Gas-driven massive black hole binaries: signatures in the nHz gravitational wave background}",
      journal = {\mnras},
         year = 2011,
        month = mar,
       volume = {411},
       number = {3},
        pages = {1467-1479},
          doi = {10.1111/j.1365-2966.2010.17782.x},
archivePrefix = {arXiv},
       eprint = {1002.0584},
 primaryClass = {astro-ph.CO},
       adsurl = {https://ui.adsabs.harvard.edu/abs/2011MNRAS.411.1467K}
}

@article{Kelley:2016gse,
    author = "Kelley, Luke Zoltan and Blecha, Laura and Hernquist, Lars",
    title = "{Massive Black Hole Binary Mergers in Dynamical Galactic Environments}",
    eprint = "1606.01900",
    archivePrefix = "arXiv",
    primaryClass = "astro-ph.HE",
    doi = "10.1093/mnras/stw2452",
    journal = "Mon. Not. Roy. Astron. Soc.",
    volume = "464",
    number = "3",
    pages = "3131--3157",
    year = "2017"
}

@article{Chen:2018znx,
    author = "Chen, Siyuan and Sesana, Alberto and Conselice, Christopher J.",
    title = "{Constraining astrophysical observables of Galaxy and Supermassive Black Hole Binary Mergers using Pulsar Timing Arrays}",
    eprint = "1810.04184",
    archivePrefix = "arXiv",
    primaryClass = "astro-ph.GA",
    doi = "10.1093/mnras/stz1722",
    journal = "Mon. Not. Roy. Astron. Soc.",
    volume = "488",
    number = "1",
    pages = "401--418",
    year = "2019"
}

@article{Bi:2023tib,
    author = "Bi, Yan-Chen and Wu, Yu-Mei and Chen, Zu-Cheng and Huang, Qing-Guo",
    title = "{Implications for the supermassive black hole binaries from the NANOGrav 15-year data set}",
    eprint = "2307.00722",
    archivePrefix = "arXiv",
    primaryClass = "astro-ph.CO",
    doi = "10.1007/s11433-023-2252-4",
    journal = "Sci. China Phys. Mech. Astron.",
    volume = "66",
    number = "12",
    pages = "120402",
    year = "2023"
}

@article{Chen:2020qlp,
    author = "Chen, Yunfeng and Yu, Qingjuan and Lu, Youjun",
    title = "{Dynamical evolution of cosmic supermassive binary black holes and their gravitational wave radiation}",
    eprint = "2005.10818",
    archivePrefix = "arXiv",
    primaryClass = "astro-ph.HE",
    doi = "10.3847/1538-4357/ab9594",
    journal = "Astrophys. J.",
    volume = "897",
    number = "1",
    pages = "86",
    year = "2020"
}

@article{Sato-Polito:2023gym,
    author = "Sato-Polito, Gabriela and Zaldarriaga, Matias and Quataert, Eliot",
    title = "{Where are the supermassive black holes measured by PTAs?}",
    eprint = "2312.06756",
    archivePrefix = "arXiv",
    primaryClass = "astro-ph.CO",
    doi = "10.1103/PhysRevD.110.063020",
    journal = "Phys. Rev. D",
    volume = "110",
    number = "6",
    pages = "063020",
    year = "2024"
}

@ARTICLE{Sato-Polito:2024lew,
       author = {{Sato-Polito}, Gabriela and {Zaldarriaga}, Matias},
        title = "{The distribution of the gravitational-wave background from supermassive black holes}",
      journal = {arXiv e-prints},
         year = 2024,
        month = jun,
          eid = {arXiv:2406.17010},
        pages = {arXiv:2406.17010},
          doi = {10.48550/arXiv.2406.17010},
archivePrefix = {arXiv},
       eprint = {2406.17010},
 primaryClass = {astro-ph.CO},
       adsurl = {https://ui.adsabs.harvard.edu/abs/2024arXiv240617010S}
}

@ARTICLE{2019PASJ...71..111I,
       author = "Izumi, Takuma and others ",
        title = "{Subaru High-z Exploration of Low-Luminosity Quasars (SHELLQs). VIII. A less biased view of the early co-evolution of black holes and host galaxies}",
      journal = {\pasj},
         year = 2019,
        month = dec,
       volume = {71},
       number = {6},
          eid = {111},
        pages = {111},
          doi = {10.1093/pasj/psz096},
archivePrefix = {arXiv},
       eprint = {1904.07345},
 primaryClass = {astro-ph.GA},
       adsurl = {https://ui.adsabs.harvard.edu/abs/2019PASJ...71..111I}
}

@ARTICLE{2020A&A...637A..84P,
       author = "Pensabene, A. and others",
        title = "{The ALMA view of the high-redshift relation between supermassive black holes and their host galaxies}",
      journal = {\aap},
         year = 2020,
        month = may,
       volume = {637},
          eid = {A84},
        pages = {A84},
          doi = {10.1051/0004-6361/201936634},
archivePrefix = {arXiv},
       eprint = {2002.00958},
 primaryClass = {astro-ph.GA},
       adsurl = {https://ui.adsabs.harvard.edu/abs/2020A&A...637A..84P}
}

@ARTICLE{2021ApJ...914...36I,
       author = "Izumi, Takuma and others",
        title = "{Subaru High-z Exploration of Low-luminosity Quasars (SHELLQs). XIII. Large-scale Feedback and Star Formation in a Low-luminosity Quasar at z = 7.07 on the Local Black Hole to Host Mass Relation}",
      journal = {\apj},
         year = 2021,
        month = jun,
       volume = {914},
       number = {1},
          eid = {36},
        pages = {36},
          doi = {10.3847/1538-4357/abf6dc},
archivePrefix = {arXiv},
       eprint = {2104.05738},
 primaryClass = {astro-ph.GA},
       adsurl = {https://ui.adsabs.harvard.edu/abs/2021ApJ...914...36I}
}

@ARTICLE{2023ApJ...959...39H,
       author = {{Harikane}, Yuichi and {Zhang}, Yechi and {Nakajima}, Kimihiko and {Ouchi}, Masami and {Isobe}, Yuki and {Ono}, Yoshiaki and {Hatano}, Shun and {Xu}, Yi and {Umeda}, Hiroya},
        title = "{A JWST/NIRSpec First Census of Broad-line AGNs at z = 4-7: Detection of 10 Faint AGNs with M $_{BH}$ {}10$^{6}$-{}10$^{8}$ M $_{{\ensuremath{\odot}}}$ and Their Host Galaxy Properties}",
      journal = {\apj},
         year = 2023,
        month = dec,
       volume = {959},
       number = {1},
          eid = {39},
        pages = {39},
          doi = {10.3847/1538-4357/ad029e},
archivePrefix = {arXiv},
       eprint = {2303.11946},
 primaryClass = {astro-ph.GA},
       adsurl = {https://ui.adsabs.harvard.edu/abs/2023ApJ...959...39H}
}

@ARTICLE{2024A&A...691A.145M,
       author = "Maiolino, Roberto and others",
        title = "{JADES: The diverse population of infant black holes at 4 < z < 11: Merging, tiny, poor, but mighty}",
      journal = {\aap},
         year = 2024,
        month = nov,
       volume = {691},
          eid = {A145},
        pages = {A145},
          doi = {10.1051/0004-6361/202347640},
archivePrefix = {arXiv},
       eprint = {2308.01230},
 primaryClass = {astro-ph.GA},
       adsurl = {https://ui.adsabs.harvard.edu/abs/2024A&A...691A.145M}
}

@ARTICLE{2024ApJ...966L..30M,
       author = {{Mezcua}, Mar and {Pacucci}, Fabio and {Suh}, Hyewon and {Siudek}, Malgorzata and {Natarajan}, Priyamvada},
        title = "{Overmassive Black Holes at Cosmic Noon: Linking the Local and the High-redshift Universe}",
      journal = {\apjl},
         year = 2024,
        month = may,
       volume = {966},
       number = {2},
          eid = {L30},
        pages = {L30},
          doi = {10.3847/2041-8213/ad3c2a},
archivePrefix = {arXiv},
       eprint = {2404.05793},
 primaryClass = {astro-ph.GA},
       adsurl = {https://ui.adsabs.harvard.edu/abs/2024ApJ...966L..30M}
}

@article{Soltan:1982vf,
    author = "Soltan, A.",
    title = "{Masses of quasars}",
    journal = "Mon. Not. Roy. Astron. Soc.",
    volume = "200",
    pages = "115--122",
    year = "1982"
}

@ARTICLE{1992MNRAS.259..725S,
       author = {{Small}, Todd A. and {Blandford}, Roger D.},
        title = "{Quasar evolution and the growth of black holes.}",
      journal = {\mnras},
         year = 1992,
        month = dec,
       volume = {259},
        pages = {725-737},
          doi = {10.1093/mnras/259.4.725},
       adsurl = {https://ui.adsabs.harvard.edu/abs/1992MNRAS.259..725S}
}

@article{Yu:2002sq,
    author = "Yu, Qing-juan and Tremaine, Scott",
    title = "{Observational constraints on growth of massive black holes}",
    eprint = "astro-ph/0203082",
    archivePrefix = "arXiv",
    doi = "10.1046/j.1365-8711.2002.05532.x",
    journal = "Mon. Not. Roy. Astron. Soc.",
    volume = "335",
    pages = "965--976",
    year = "2002"
}

@ARTICLE{2012AdAst2012E...7K,
       author = {{Kelly}, Brandon C. and {Merloni}, Andrea},
        title = "{Mass Functions of Supermassive Black Holes across Cosmic Time}",
      journal = {Advances in Astronomy},
         year = 2012,
        month = jan,
       volume = {2012},
          eid = {970858},
        pages = {970858},
          doi = {10.1155/2012/970858},
archivePrefix = {arXiv},
       eprint = {1112.1430},
 primaryClass = {astro-ph.CO},
       adsurl = {https://ui.adsabs.harvard.edu/abs/2012AdAst2012E...7K}
}

@article{Ueda:2014tma,
    author = {Ueda, Yoshihiro and Akiyama, Masayuki and Hasinger, G\"unther and Miyaji, Takamitsu and Watson, Michael G.},
    title = "{Toward the Standard Population Synthesis Model of the X-Ray Background: Evolution of X-Ray Luminosity and Absorption Functions of Active Galactic Nuclei Including Compton-Thick Populations}",
    eprint = "1402.1836",
    archivePrefix = "arXiv",
    primaryClass = "astro-ph.CO",
    doi = "10.1088/0004-637X/786/2/104",
    journal = "Astrophys. J.",
    volume = "786",
    pages = "104",
    year = "2014"
}

@article{Tucci:2016tyc,
    author = "Tucci, Marco and Volonteri, Marta",
    title = "{Constraining supermassive black hole evolution through the continuity equation}",
    eprint = "1603.00823",
    archivePrefix = "arXiv",
    primaryClass = "astro-ph.GA",
    doi = "10.1051/0004-6361/201628419",
    journal = "Astron. Astrophys.",
    volume = "600",
    pages = "A64",
    year = "2017"
}

@article{Shen:2020obl,
    author = "Shen, Xuejian and Hopkins, Philip F. and Faucher-Gigu\`ere, Claude-Andr\'e and Alexander, D. M. and Richards, Gordon T. and Ross, Nicholas P. and Hickox, R. C.",
    title = "{The bolometric quasar luminosity function at z = 0\textendash{}7}",
    eprint = "2001.02696",
    archivePrefix = "arXiv",
    primaryClass = "astro-ph.GA",
    doi = "10.1093/mnras/staa1381",
    journal = "Mon. Not. Roy. Astron. Soc.",
    volume = "495",
    number = "3",
    pages = "3252--3275",
    year = "2020"
}

@article{Goulding:2019hnn,
    author = "Goulding, Andy D. and Pardo, Kris and Greene, Jenny E. and Mingarelli, Chiara M. F. and Nyland, Kristina and Strauss, Michael A.",
    title = "{Discovery of a Close-separation Binary Quasar at the Heart of a z \ensuremath{\sim} 0.2 Merging Galaxy and Its Implications for Low-frequency Gravitational Waves}",
    eprint = "1907.03757",
    archivePrefix = "arXiv",
    primaryClass = "astro-ph.GA",
    doi = "10.3847/2041-8213/ab2a14",
    journal = "Astrophys. J. Lett.",
    volume = "879",
    number = "2",
    pages = "L21",
    year = "2019"
}

@article{Casey-Clyde:2021xro,
    author = "Casey-Clyde, J. Andrew and Mingarelli, Chiara M. F. and Greene, Jenny E. and Pardo, Kris and Na\~nez, Morgan and Goulding, Andy D.",
    title = "{A Quasar-based Supermassive Black Hole Binary Population Model: Implications for the Gravitational Wave Background}",
    eprint = "2107.11390",
    archivePrefix = "arXiv",
    primaryClass = "astro-ph.HE",
    doi = "10.3847/1538-4357/ac32de",
    journal = "Astrophys. J.",
    volume = "924",
    number = "2",
    pages = "93",
    year = "2022"
}

@article{Casey-Clyde:2024hwg,
    author = "Casey-Clyde, J. Andrew and Mingarelli, Chiara M. F. and Greene, Jenny E. and Goulding, Andy D. and Chen, Siyuan and Trump, Jonathan R.",
    title = "{Quasars can Signpost Supermassive Black Hole Binaries}",
    eprint = "2405.19406",
    archivePrefix = "arXiv",
    primaryClass = "astro-ph.HE",
    month = "5",
    year = "2024"
}

@article{Kis-Toth:2024gkm,
    author = "Kis-T\'oth, \'Agnes and Haiman, Zolt\'an and Frei, Zsolt",
    title = "{Can quasars, triggered by mergers, account for NANOGrav's stochastic gravitational wave background?}",
    eprint = "2412.12726",
    archivePrefix = "arXiv",
    primaryClass = "astro-ph.CO",
    month = "12",
    year = "2024"
}

@article{Begelman:1980vb,
    author = "Begelman, M. C. and Blandford, R. D. and Rees, M. J.",
    title = "{Massive black hole binaries in active galactic nuclei}",
    doi = "10.1038/287307a0",
    journal = "Nature",
    volume = "287",
    pages = "307--309",
    year = "1980"
}

@article{Kelley:2017lek,
    author = "Kelley, Luke Zoltan and Blecha, Laura and Hernquist, Lars and Sesana, Alberto and Taylor, Stephen R.",
    title = "{The Gravitational Wave Background from Massive Black Hole Binaries in Illustris: spectral features and time to detection with pulsar timing arrays}",
    eprint = "1702.02180",
    archivePrefix = "arXiv",
    primaryClass = "astro-ph.HE",
    doi = "10.1093/mnras/stx1638",
    journal = "Mon. Not. Roy. Astron. Soc.",
    volume = "471",
    number = "4",
    pages = "4508--4526",
    year = "2017"
}

@article{Sesana:2012ak,
    author = "Sesana, A.",
    title = "{Systematic investigation of the expected gravitational wave signal from supermassive black hole binaries in the pulsar timing band}",
    eprint = "1211.5375",
    archivePrefix = "arXiv",
    primaryClass = "astro-ph.CO",
    doi = "10.1093/mnrasl/slt034",
    journal = "Mon. Not. Roy. Astron. Soc.",
    volume = "433",
    pages = "1",
    year = "2013"
}

@ARTICLE{2019astro2020T.504K,
       author = "Koss, Michael and others",
        title = "{Black Hole Growth in Mergers and Dual AGN}",
      journal = {Astro2020: Decadal Survey on Astronomy and Astrophysics},
         year = 2019,
        month = may,
       volume = {2020},
        pages = {504},
          doi = {10.48550/arXiv.1903.06720},
archivePrefix = {arXiv},
       eprint = {1903.06720},
 primaryClass = {astro-ph.GA},
       adsurl = {https://ui.adsabs.harvard.edu/abs/2019astro2020T.504K}
}

@article{Sesana:2016yky,
    author = "Sesana, Alberto and Shankar, Francesco and Bernardi, Mariangela and Sheth, Ravi K.",
    title = "{Selection bias in dynamically measured supermassive black hole samples: consequences for pulsar timing arrays}",
    eprint = "1603.09348",
    archivePrefix = "arXiv",
    primaryClass = "astro-ph.GA",
    doi = "10.1093/mnrasl/slw139",
    journal = "Mon. Not. Roy. Astron. Soc.",
    volume = "463",
    number = "1",
    pages = "L6--L11",
    year = "2016"
}

@article{Sudou:2003hv,
    author = "Sudou, Hiroshi and Iguchi, Satoru and Murata, Yasuhiro and Taniguchi, Yoshiaki",
    title = "{Orbital motion in the radio galaxy 3c 66b: evidence for a supermassive black hole binary}",
    eprint = "astro-ph/0306103",
    archivePrefix = "arXiv",
    doi = "10.1126/science.1082817",
    journal = "Science",
    volume = "300",
    pages = "1263",
    year = "2003"
}

@article{Bon:2016jtk,
    author = "Bon, E. and others",
    title = "{Evidence for periodicity in 43 year-long monitoring of NGC 5548}",
    eprint = "1606.04606",
    archivePrefix = "arXiv",
    primaryClass = "astro-ph.HE",
    doi = "10.3847/0067-0049/225/2/29",
    journal = "Astrophys. J. Suppl.",
    volume = "225",
    number = "2",
    pages = "29",
    year = "2016"
}

@article{Runnoe:2017oxn,
    author = "Runnoe, Jessie C. and others",
    title = "{A large systematic search for close supermassive binary and rapidly recoiling black holes \textendash{} III. Radial velocity variations}",
    eprint = "1702.05465",
    archivePrefix = "arXiv",
    primaryClass = "astro-ph.GA",
    doi = "10.1093/mnras/stx452",
    journal = "Mon. Not. Roy. Astron. Soc.",
    volume = "468",
    number = "2",
    pages = "1683--1702",
    year = "2017"
}

@article{Guo:2018xum,
    author = "Guo, Hengxiao and Liu, Xin and Shen, Yue and Loeb, Abraham and Monroe, TalaWanda and Prochaska, Javier Xavier",
    title = "{Constraining sub-parsec binary supermassive black holes in quasars with multi-epoch spectroscopy \textendash{} III. Candidates from continued radial velocity tests}",
    eprint = "1809.04610",
    archivePrefix = "arXiv",
    primaryClass = "astro-ph.GA",
    doi = "10.1093/mnras/sty2920",
    journal = "Mon. Not. Roy. Astron. Soc.",
    volume = "482",
    number = "3",
    pages = "3288--3307",
    year = "2019"
}

@article{Adhikari:2023uju,
    author = "Adhikari, Sagar and Penil, Pablo and Westernacher-Schneider, John Ryan and Dominguez, Alberto and Ajello, Marco and Buson, Sara and Rico, Alba and Zrake, Jonathan",
    title = "{Constraining the PG 1553+113 Binary Hypothesis: Interpreting Hints of a New, 22 yr Period}",
    eprint = "2307.11696",
    archivePrefix = "arXiv",
    primaryClass = "astro-ph.HE",
    doi = "10.3847/1538-4357/ad310a",
    journal = "Astrophys. J.",
    volume = "965",
    number = "2",
    pages = "124",
    year = "2024"
}

@article{Zhou:2004uq,
    author = "Zhou, Hong-Yan and Wang, Ting-Gui and Zhang, Xue-Guang and Dong, Xiao-Bo and Li, Cheng",
    title = "{Obscured binary quasar cores in SDSS J104807.74+005543.5?}",
    eprint = "astro-ph/0411167",
    archivePrefix = "arXiv",
    doi = "10.1086/383310",
    journal = "Astrophys. J. Lett.",
    volume = "604",
    pages = "L33--L36",
    year = "2004"
}

@ARTICLE{2012ApJ...759..118B,
       author = "Bon, E. and others ",
        title = "{The First Spectroscopically Resolved Sub-parsec Orbit of a Supermassive Binary Black Hole}",
      journal = {\apj},
         year = 2012,
        month = nov,
       volume = {759},
       number = {2},
          eid = {118},
        pages = {118},
          doi = {10.1088/0004-637X/759/2/118},
archivePrefix = {arXiv},
       eprint = {1209.4524},
 primaryClass = {astro-ph.HE},
       adsurl = {https://ui.adsabs.harvard.edu/abs/2012ApJ...759..118B}
}

@ARTICLE{2014ApJ...789..140L,
       author = {{Liu}, Xin and {Shen}, Yue and {Bian}, Fuyan and {Loeb}, Abraham and {Tremaine}, Scott},
        title = "{Constraining Sub-parsec Binary Supermassive Black Holes in Quasars with Multi-epoch Spectroscopy. II. The Population with Kinematically Offset Broad Balmer Emission Lines}",
      journal = {\apj},
         year = 2014,
        month = jul,
       volume = {789},
       number = {2},
          eid = {140},
        pages = {140},
          doi = {10.1088/0004-637X/789/2/140},
archivePrefix = {arXiv},
       eprint = {1312.6694},
 primaryClass = {astro-ph.CO},
       adsurl = {https://ui.adsabs.harvard.edu/abs/2014ApJ...789..140L}
}

@article{Rodriguez:2006th,
    author = "Rodriguez, C. and Taylor, Greg B. and Zavala, R. T. and Peck, A. B. and Pollack, L. K. and Romani, R. W.",
    title = "{A Compact Supermassive Binary Black Hole System}",
    eprint = "astro-ph/0604042",
    archivePrefix = "arXiv",
    doi = "10.1086/504825",
    journal = "Astrophys. J.",
    volume = "646",
    pages = "49--60",
    year = "2006"
}

@article{Deane:2014jqa,
    author = "Deane, R. P. and others",
    title = "{A close-pair binary in a distant triple supermassive black-hole system}",
    eprint = "1406.6365",
    archivePrefix = "arXiv",
    primaryClass = "astro-ph.GA",
    doi = "10.1038/nature13454",
    journal = "Nature",
    volume = "511",
    pages = "57",
    year = "2014"
}

@ARTICLE{2017NatAs...1..727K,
       author = {{Kharb}, P. and {Lal}, D.~V. and {Merritt}, D.},
        title = "{A candidate sub-parsec binary black hole in the Seyfert galaxy NGC 7674}",
      journal = {Nature Astronomy},
         year = 2017,
        month = sep,
       volume = {1},
        pages = {727-733},
          doi = {10.1038/s41550-017-0256-4},
archivePrefix = {arXiv},
       eprint = {1709.06258},
 primaryClass = {astro-ph.GA},
       adsurl = {https://ui.adsabs.harvard.edu/abs/2017NatAs...1..727K}
}

@article{Goulding:2023gqa,
    author = "Goulding, Andy D. and others",
    title = "{UNCOVER: The Growth of the First Massive Black Holes from JWST/NIRSpec\textemdash{}Spectroscopic Redshift Confirmation of an X-Ray Luminous AGN at z = 10.1}",
    eprint = "2308.02750",
    archivePrefix = "arXiv",
    primaryClass = "astro-ph.GA",
    doi = "10.3847/2041-8213/acf7c5",
    journal = "Astrophys. J. Lett.",
    volume = "955",
    number = "1",
    pages = "L24",
    year = "2023"
}

@article{Kovacs:2024zfh,
    author = "Kovacs, Orsolya E. and others",
    title = "{A Candidate Supermassive Black Hole in a Gravitationally Lensed Galaxy at Z \ensuremath{\approx} 10}",
    eprint = "2403.14745",
    archivePrefix = "arXiv",
    primaryClass = "astro-ph.GA",
    doi = "10.3847/2041-8213/ad391f",
    journal = "Astrophys. J. Lett.",
    volume = "965",
    number = "2",
    pages = "L21",
    year = "2024"
}

@ARTICLE{2023arXiv231003067P,
       author = "Perna, Michele and others",
      journal = {arXiv e-prints},
         year = 2023,
        month = oct,
          eid = {arXiv:2310.03067},
        pages = {arXiv:2310.03067},
          doi = {10.48550/arXiv.2310.03067},
archivePrefix = {arXiv},
       eprint = {2310.03067},
 primaryClass = {astro-ph.GA},
       adsurl = {https://ui.adsabs.harvard.edu/abs/2023arXiv231003067P}
}

@ARTICLE{2024MNRAS.531..355U,
       author = "{{\"U}bler}, Hannah and {Maiolino}, Roberto and {P{\'e}rez-Gonz{\'a}lez} and others",
        title = "{GA-NIFS: JWST discovers an offset AGN 740 million years after the big bang}",
      journal = {\mnras},
         year = 2024,
        month = jun,
       volume = {531},
       number = {1},
        pages = {355-365},
          doi = {10.1093/mnras/stae943},
archivePrefix = {arXiv},
       eprint = {2312.03589},
 primaryClass = {astro-ph.GA},
       adsurl = {https://ui.adsabs.harvard.edu/abs/2024MNRAS.531..355U}
}

@article{Ellis:2024wdh,
    author = {Ellis, John and Fairbairn, Malcolm and H\"utsi, Gert and Urrutia, Juan and Vaskonen, Ville and Veerm\"ae, Hardi},
    title = "{Consistency of JWST Black Hole Observations with NANOGrav Gravitational Wave Measurements}",
    eprint = "2403.19650",
    archivePrefix = "arXiv",
    primaryClass = "astro-ph.CO",
    reportNumber = "KCL-PH-TH/2024-16, CERN-TH-2024-038, AION-REPORT/2024-03",
    doi = "10.1051/0004-6361/202450846",
    journal = "Astron. Astrophys.",
    volume = "691",
    pages = "A270",
    year = "2024"
}

@article{Padmanabhan:2024nvv,
    author = "Padmanabhan, Hamsa and Loeb, Abraham",
    title = "{Constraints on supermassive black hole binaries from JWST and NANOGrav}",
    eprint = "2401.04161",
    archivePrefix = "arXiv",
    primaryClass = "astro-ph.HE",
    doi = "10.1051/0004-6361/202449185",
    journal = "Astron. Astrophys.",
    volume = "684",
    pages = "L15",
    year = "2024"
}

@ARTICLE{2011ApJ...737..101L,
       author = {{Liu}, Xin and {Shen}, Yue and {Strauss}, Michael A. and {Hao}, Lei},
        title = "{Active Galactic Nucleus Pairs from the Sloan Digital Sky Survey. I. The Frequency on \raisebox{-0.5ex}\textasciitilde5-100 kpc Scales}",
      journal = {\apj},
         year = 2011,
        month = aug,
       volume = {737},
       number = {2},
          eid = {101},
        pages = {101},
          doi = {10.1088/0004-637X/737/2/101},
archivePrefix = {arXiv},
       eprint = {1104.0950},
 primaryClass = {astro-ph.CO},
       adsurl = {https://ui.adsabs.harvard.edu/abs/2011ApJ...737..101L}
}

@ARTICLE{2012ApJ...746L..22K,
       author = {{Koss}, Michael and {Mushotzky}, Richard and {Treister}, Ezequiel and {Veilleux}, Sylvain and {Vasudevan}, Ranjan and {Trippe}, Margaret},
        title = "{Understanding Dual Active Galactic Nucleus Activation in the nearby Universe}",
      journal = {\apjl},
         year = 2012,
        month = feb,
       volume = {746},
       number = {2},
          eid = {L22},
        pages = {L22},
          doi = {10.1088/2041-8205/746/2/L22},
archivePrefix = {arXiv},
       eprint = {1201.2944},
 primaryClass = {astro-ph.HE},
       adsurl = {https://ui.adsabs.harvard.edu/abs/2012ApJ...746L..22K}
}

@ARTICLE{2020ApJ...899..154S,
       author = "Silverman, John D. and others",
        title = "{Dual Supermassive Black Holes at Close Separation Revealed by the Hyper Suprime-Cam Subaru Strategic Program}",
      journal = {\apj},
         year = 2020,
        month = aug,
       volume = {899},
       number = {2},
          eid = {154},
        pages = {154},
          doi = {10.3847/1538-4357/aba4a3},
archivePrefix = {arXiv},
       eprint = {2007.05581},
 primaryClass = {astro-ph.GA},
       adsurl = {https://ui.adsabs.harvard.edu/abs/2020ApJ...899..154S}
}

@article{Shen:2022cmp,
    author = "Shen, Yue and others",
    title = "{Statistics of Galactic-scale Quasar Pairs at Cosmic Noon}",
    eprint = "2208.04979",
    archivePrefix = "arXiv",
    primaryClass = "astro-ph.GA",
    doi = "10.3847/1538-4357/aca662",
    journal = "Astrophys. J.",
    volume = "943",
    number = "1",
    pages = "38",
    year = "2023"
}

@ARTICLE{2024arXiv240514980L,
       author = {{Li}, Junyao and {Zhuang}, Ming-Yang and {Shen}, Yue and {Volonteri}, Marta and {Chen}, Nianyi and {Di Matteo}, Tiziana},
        title = "{Active Galactic Nuclei and Host Galaxies in COSMOS-Web. II. First Look at the Kpc-scale Dual and Offset AGN Population}",
      journal = {arXiv e-prints},
         year = 2024,
        month = may,
          eid = {arXiv:2405.14980},
        pages = {arXiv:2405.14980},
          doi = {10.48550/arXiv.2405.14980},
archivePrefix = {arXiv},
       eprint = {2405.14980},
 primaryClass = {astro-ph.GA},
       adsurl = {https://ui.adsabs.harvard.edu/abs/2024arXiv240514980L}
}

@article{Planck:2018vyg,
    author = "Aghanim, N. and others",
    collaboration = "Planck",
    title = "{Planck 2018 results. VI. Cosmological parameters}",
    eprint = "1807.06209",
    archivePrefix = "arXiv",
    primaryClass = "astro-ph.CO",
    doi = "10.1051/0004-6361/201833910",
    journal = "Astron. Astrophys.",
    volume = "641",
    pages = "A6",
    year = "2020",
    note = "[Erratum: Astron.Astrophys. 652, C4 (2021)]"
}

@ARTICLE{2001astro.ph..8028P,
       author = {{Phinney}, E.~S.},
        title = "{A Practical Theorem on Gravitational Wave Backgrounds}",
      journal = {arXiv e-prints},
         year = 2001,
        month = aug,
          eid = {astro-ph/0108028},
        pages = {astro-ph/0108028},
          doi = {10.48550/arXiv.astro-ph/0108028},
archivePrefix = {arXiv},
       eprint = {astro-ph/0108028},
 primaryClass = {astro-ph},
       adsurl = {https://ui.adsabs.harvard.edu/abs/2001astro.ph..8028P}
}

@article{Chen:2016zyo,
    author = "Chen, Siyuan and Sesana, Alberto and Del Pozzo, Walter",
    title = "{Efficient computation of the gravitational wave spectrum emitted by eccentric massive black hole binaries in stellar environments}",
    eprint = "1612.00455",
    archivePrefix = "arXiv",
    primaryClass = "astro-ph.CO",
    doi = "10.1093/mnras/stx1093",
    journal = "Mon. Not. Roy. Astron. Soc.",
    volume = "470",
    number = "2",
    pages = "1738--1749",
    year = "2017"
}

@article{Aird:2015fya,
    author = "Aird, James and Coil, Alison L. and Georgakakis, Antonis and Nandra, Kirpal and Barro, Guillermo and P\'erez-Gonz\'alez, Pablo G.",
    title = "{The evolution of the X-ray luminosity functions of unabsorbed and absorbed AGNs out to z\ensuremath{\sim} 5}",
    eprint = "1503.01120",
    archivePrefix = "arXiv",
    primaryClass = "astro-ph.HE",
    doi = "10.1093/mnras/stv1062",
    journal = "Mon. Not. Roy. Astron. Soc.",
    volume = "451",
    number = "2",
    pages = "1892--1927",
    year = "2015"
}

@article{PhysRev.136.B1224,
  title = {Gravitational Radiation and the Motion of Two Point Masses},
  author = {Peters, P. C.},
  journal = {Phys. Rev.},
  volume = {136},
  issue = {4B},
  pages = {B1224--B1232},
  numpages = {0},
  year = {1964},
  month = {Nov},
  publisher = {American Physical Society},
  doi = {10.1103/PhysRev.136.B1224},
  url = {https://link.aps.org/doi/10.1103/PhysRev.136.B1224}
}

@article{Goulding:2017mul,
    author = "Goulding, Andy D. and Greene, Jenny E. and Bezanson, Rachel and Greco, Johnny and Johnson, Sean and Leauthaud, Alexie and Matsuoka, Yoshiki and Medezinski, Elinor and Price-Whelan, Adrian M.",
    title = "{Galaxy interactions trigger rapid black hole growth: An unprecedented view from the Hyper Suprime-Cam survey}",
    eprint = "1706.07436",
    archivePrefix = "arXiv",
    primaryClass = "astro-ph.GA",
    doi = "10.1093/pasj/psx135",
    journal = "Publ. Astron. Soc. Jap.",
    volume = "70",
    number = "SP1",
    pages = "Publications of the Astronomical Society of Japan, Volume 70, Issue SP1, January 2018, S37, https://doi.org/10.1093/pasj/psx135",
    year = "2018"
}

@article{Conselice:2014hqa,
    author = "Conselice, Christopher J. and Bluck, Asa F. L. and Mortlock, Alice and Palamara, David and Benson, Andrew J.",
    title = "{Galaxy formation as a cosmological tool {\textendash} I. The galaxy merger history as a measure of cosmological parameters}",
    eprint = "1407.3811",
    archivePrefix = "arXiv",
    primaryClass = "astro-ph.GA",
    doi = "10.1093/mnras/stu1385",
    journal = "Mon. Not. Roy. Astron. Soc.",
    volume = "444",
    number = "2",
    pages = "1125--1143",
    year = "2014"
}

@ARTICLE{2014AJ....148..137L,
       author = {{Lackner}, C.~N. and {Silverman}, J.~D. and {Salvato}, M. and {Kampczyk}, P. and {Kartaltepe}, J.~S. and {Sanders}, D. and {Capak}, P. and {Civano}, F. and {Halliday}, C. and {Ilbert}, O. and {Jahnke}, K. and {Koekemoer}, A.~M. and {Lee}, N. and {Le F{\`e}vre}, O. and {Liu}, C.~T. and {Scoville}, N. and {Sheth}, K. and {Toft}, S.},
        title = "{Late-Stage Galaxy Mergers in Cosmos to z {\ensuremath{\sim}} 1}",
      journal = {\aj},
         year = 2014,
        month = dec,
       volume = {148},
       number = {6},
          eid = {137},
        pages = {137},
          doi = {10.1088/0004-6256/148/6/137},
archivePrefix = {arXiv},
       eprint = {1406.2327},
 primaryClass = {astro-ph.GA},
       adsurl = {https://ui.adsabs.harvard.edu/abs/2014AJ....148..137L}
}

@article{Quinlan:1996vp,
    author = "Quinlan, Gerald D.",
    title = "{The dynamical evolution of massive black hole binaries - I. hardening in a fixed stellar background}",
    eprint = "astro-ph/9601092",
    archivePrefix = "arXiv",
    reportNumber = "RUTGERS-ASTROPHYSICS-PREPRINT-SERIES-NO-187",
    doi = "10.1016/S1384-1076(96)00003-6",
    journal = "New Astron.",
    volume = "1",
    pages = "35--56",
    year = "1996"
}

@BOOK{2008gady.book.....B,
       author = {{Binney}, James and {Tremaine}, Scott},
        title = "{Galactic Dynamics: Second Edition}",
         year = 2008,
       adsurl = {https://ui.adsabs.harvard.edu/abs/2008gady.book.....B}
}

@article{Dosopoulou:2016hbg,
    author = "Dosopoulou, Fani and Antonini, Fabio",
    title = "{Dynamical friction and the evolution of Supermassive Black hole Binaries: the final hundred-parsec problem}",
    eprint = "1611.06573",
    archivePrefix = "arXiv",
    primaryClass = "astro-ph.GA",
    doi = "10.3847/1538-4357/aa6b58",
    journal = "Astrophys. J.",
    volume = "840",
    number = "1",
    pages = "31",
    year = "2017"
}

@article{DOrazio:2017dyb,
    author = "D'Orazio, Daniel J. and Loeb, Abraham",
    title = "{Repeated Imaging of Massive Black Hole Binary Orbits with Millimeter Interferometry: Measuring Black Hole Masses and the Hubble Constant}",
    eprint = "1712.02362",
    archivePrefix = "arXiv",
    primaryClass = "astro-ph.HE",
    doi = "10.3847/1538-4357/aad413",
    journal = "Astrophys. J.",
    volume = "863",
    number = "2",
    pages = "185",
    year = "2018"
}

@article{Loeb:2009rv,
    author = "Loeb, Abraham",
    title = "{Electromagnetic signature of supermassive black hole binaries that enter their gravitational-wave induced inspiral}",
    eprint = "0909.0261",
    archivePrefix = "arXiv",
    primaryClass = "astro-ph.CO",
    doi = "10.1103/PhysRevD.81.047503",
    journal = "Phys. Rev. D",
    volume = "81",
    pages = "047503",
    year = "2010"
}

@article{Sesana:2015haa,
    author = "Sesana, Alberto and Khan, Fazeel Mahmood",
    title = "{Scattering experiments meet N-body \textendash{} I. A practical recipe for the evolution of massive black hole binaries in stellar environments}",
    eprint = "1505.02062",
    archivePrefix = "arXiv",
    primaryClass = "astro-ph.GA",
    doi = "10.1093/mnrasl/slv131",
    journal = "Mon. Not. Roy. Astron. Soc.",
    volume = "454",
    number = "1",
    pages = "L66--L70",
    year = "2015"
}

@article{Zhao:2023kff,
    author = "Zhao, Shan-Shan and Jiang, Wu and Lu, Ru-Sen and Huang, Lei and Shen, Zhi-Qiang",
    title = "{How Many Supermassive Black Hole Binaries Are Detectable through Tracking Relative Motions by (Sub)millimeter Very Long Baseline Interferometry?}",
    eprint = "2311.11589",
    archivePrefix = "arXiv",
    primaryClass = "astro-ph.GA",
    doi = "10.3847/1538-4357/ad0da1",
    journal = "Astrophys. J.",
    volume = "961",
    number = "1",
    pages = "20",
    year = "2024"
}

@article{Tremaine:2002js,
    author = "Tremaine, Scott and others",
    title = "{The slope of the black hole mass versus velocity dispersion correlation}",
    eprint = "astro-ph/0203468",
    archivePrefix = "arXiv",
    doi = "10.1086/341002",
    journal = "Astrophys. J.",
    volume = "574",
    pages = "740--753",
    year = "2002"
}

@article{Cohen:1997tx,
    author = "Cohen, Judith G. and Ryzhov, Anton",
    title = "{The dynamics of the m87 globular cluster system}",
    eprint = "astro-ph/9704051",
    archivePrefix = "arXiv",
    doi = "10.1086/304518",
    journal = "Astrophys. J.",
    volume = "486",
    pages = "230",
    year = "1997"
}

@ARTICLE{2018MNRAS.476.2308W,
       author = {{Weigel}, Anna K. and {Schawinski}, Kevin and {Treister}, Ezequiel and {Trakhtenbrot}, Benny and {Sanders}, David B.},
        title = "{The fraction of AGNs in major merger galaxies and its luminosity dependence}",
      journal = {\mnras},
         year = 2018,
        month = may,
       volume = {476},
       number = {2},
        pages = {2308-2317},
          doi = {10.1093/mnras/sty383},
archivePrefix = {arXiv},
       eprint = {1802.04277},
 primaryClass = {astro-ph.GA},
       adsurl = {https://ui.adsabs.harvard.edu/abs/2018MNRAS.476.2308W}
}

@article{Lapi:2025wxt,
    author = "Lapi, Andrea and Shankar, Francesco and Bosi, Michele and Roberts, Daniel and Fu, Hao and Varadarajan, Karthik M. and Boco, Lumen",
    title = "{Semi-empirical Modeling of Supermassive Black Hole Evolution: Highlighting a possible tension between Demographics and Gravitational Wave Background}",
    eprint = "2507.15436",
    archivePrefix = "arXiv",
    primaryClass = "astro-ph.CO",
    month = "7",
    year = "2025"
}

@article{Sato-Polito:2025dqw,
    author = "Sato-Polito, Gabriela and Zaldarriaga, Matias",
    title = "{Uncertainties in the supermassive black hole abundance and implications for the GW background}",
    eprint = "2509.08041",
    archivePrefix = "arXiv",
    primaryClass = "astro-ph.GA",
    month = "9",
    year = "2025"
}

@inproceedings{Martini:2003ek,
    author = "Martini, Paul",
    title = "{QSO lifetimes}",
    booktitle = "{Carnegie Observatories Centennial Symposium. 1. Coevolution of Black Holes and Galaxies}",
    eprint = "astro-ph/0304009",
    archivePrefix = "arXiv",
    month = "4",
    year = "2003"
}

@article{Lai:2022ylu,
    author = "Lai, Dong and Mu{\~n}oz, Diego J.",
    title = "{Circumbinary Accretion: From Binary Stars to Massive Binary Black Holes}",
    eprint = "2211.00028",
    archivePrefix = "arXiv",
    primaryClass = "astro-ph.HE",
    doi = "10.1146/annurev-astro-052622-022933",
    journal = "Ann. Rev. Astron. Astrophys.",
    volume = "61",
    pages = "517--560",
    year = "2023"
}

@article{Ellis:2023dgf,
    author = {Ellis, John and Fairbairn, Malcolm and H{\"u}tsi, Gert and Raidal, Juhan and Urrutia, Juan and Vaskonen, Ville and Veerm{\"a}e, Hardi},
    title = "{Gravitational waves from supermassive black hole binaries in light of the NANOGrav 15-year data}",
    eprint = "2306.17021",
    archivePrefix = "arXiv",
    primaryClass = "astro-ph.CO",
    reportNumber = "KCL-PH-TH/2023-37, CERN-TH-2023-120, AION-REPORT/2023-06",
    doi = "10.1103/PhysRevD.109.L021302",
    journal = "Phys. Rev. D",
    volume = "109",
    number = "2",
    pages = "L021302",
    year = "2024"
}

@article{Raidal:2024odr,
    author = {Raidal, Juhan and Urrutia, Juan and Vaskonen, Ville and Veerm{\"a}e, Hardi},
    title = "{Eccentricity effects on the supermassive black hole gravitational wave background}",
    eprint = "2406.05125",
    archivePrefix = "arXiv",
    primaryClass = "astro-ph.CO",
    doi = "10.1051/0004-6361/202451345",
    journal = "Astron. Astrophys.",
    volume = "691",
    pages = "A212",
    year = "2024"
}

@ARTICLE{2017ApJ...835...27A,
       author = {{Azadi}, Mojegan and {Coil}, Alison L. and {Aird}, James and {Reddy}, Naveen and {Shapley}, Alice and {Freeman}, William R. and {Kriek}, Mariska and {Leung}, Gene C.~K. and {Mobasher}, Bahram and {Price}, Sedona H. and {Sanders}, Ryan L. and {Shivaei}, Irene and {Siana}, Brian},
        title = "{The MOSDEF Survey: AGN Multi-wavelength Identification, Selection Biases, and Host Galaxy Properties}",
      journal = {\apj},
         year = 2017,
        month = jan,
       volume = {835},
       number = {1},
          eid = {27},
        pages = {27},
          doi = {10.3847/1538-4357/835/1/27},
archivePrefix = {arXiv},
       eprint = {1608.05890},
 primaryClass = {astro-ph.GA},
       adsurl = {https://ui.adsabs.harvard.edu/abs/2017ApJ...835...27A}
}

@ARTICLE{2020ApJ...888...78S,
       author = {{Stemo}, Aaron and {Comerford}, Julia M. and {Barrows}, R. Scott and {Stern}, Daniel and {Assef}, Roberto J. and {Griffith}, Roger L.},
        title = "{A Catalog of AGN Host Galaxies Observed with HST/ACS: Correlations between Star Formation and AGN Activity}",
      journal = {\apj},
         year = 2020,
        month = jan,
       volume = {888},
       number = {2},
          eid = {78},
        pages = {78},
          doi = {10.3847/1538-4357/ab5f66},
archivePrefix = {arXiv},
       eprint = {1911.07864},
 primaryClass = {astro-ph.GA},
       adsurl = {https://ui.adsabs.harvard.edu/abs/2020ApJ...888...78S}
}

@article{Capelo:2016yrq,
    author = "Capelo, Pedro R. and Dotti, Massimo and Volonteri, Marta and Mayer, Lucio and Bellovary, Jillian M. and Shen, Sijing",
    title = "{A survey of dual active galactic nuclei in simulations of galaxy mergers: frequency and properties}",
    eprint = "1611.09244",
    archivePrefix = "arXiv",
    primaryClass = "astro-ph.GA",
    doi = "10.1093/mnras/stx1067",
    journal = "Mon. Not. Roy. Astron. Soc.",
    volume = "469",
    number = "4",
    pages = "4437--4454",
    year = "2017"
}

@article{Hopkins:2005fb,
    author = "Hopkins, Philip F. and Hernquist, Lars and Cox, Thomas J. and Di Matteo, Tiziana and Robertson, Brant and Springel, Volker",
    title = "{A Unified, merger-driven model for the origin of starbursts, quasars, the cosmic x-ray background, supermassive black holes and galaxy spheroids}",
    eprint = "astro-ph/0506398",
    archivePrefix = "arXiv",
    doi = "10.1086/499298",
    journal = "Astrophys. J. Suppl.",
    volume = "163",
    pages = "1--49",
    year = "2006"
}

@ARTICLE{2010MNRAS.406.1959S,
       author = "{Shankar}, Francesco and {Weinberg}, David H. and {Shen}, Yue",
        title = "{Constraints on black hole duty cycles and the black hole-halo relation from SDSS quasar clustering}",
      journal = {\mnras},
         year = 2010,
        month = aug,
       volume = {406},
       number = {3},
        pages = {1959-1966},
          doi = {10.1111/j.1365-2966.2010.16801.x},
archivePrefix = {arXiv},
       eprint = {1004.1173},
 primaryClass = {astro-ph.CO},
       adsurl = {https://ui.adsabs.harvard.edu/abs/2010MNRAS.406.1959S}
}

@ARTICLE{2013MNRAS.428..421S,
       author = "{Shankar}, Francesco and {Weinberg}, David H. and {Miralda-Escud{\'e}}, Jordi",
        title = "{Accretion-driven evolution of black holes: Eddington ratios, duty cycles and active galaxy fractions}",
      journal = {\mnras},
         year = 2013,
        month = jan,
       volume = {428},
       number = {1},
        pages = {421-446},
          doi = {10.1093/mnras/sts026},
archivePrefix = {arXiv},
       eprint = {1111.3574},
 primaryClass = {astro-ph.CO},
       adsurl = {https://ui.adsabs.harvard.edu/abs/2013MNRAS.428..421S}
}

@ARTICLE{2017ApJ...845..145W,
       author = "Weigel, Anna K. and others",
        title = "{Galaxy Zoo: Major Galaxy Mergers Are Not a Significant Quenching Pathway}",
      journal = {\apj},
         year = 2017,
        month = aug,
       volume = {845},
       number = {2},
          eid = {145},
        pages = {145},
          doi = {10.3847/1538-4357/aa8097},
archivePrefix = {arXiv},
       eprint = {1708.00866},
 primaryClass = {astro-ph.GA},
       adsurl = {https://ui.adsabs.harvard.edu/abs/2017ApJ...845..145W}
}

@ARTICLE{2012ApJ...758L..39T,
       author = {{Treister}, E. and {Schawinski}, K. and {Urry}, C.~M. and {Simmons}, B.~D.},
        title = "{Major Galaxy Mergers Only Trigger the Most Luminous Active Galactic Nuclei}",
      journal = {\apjl},
         year = 2012,
        month = oct,
       volume = {758},
       number = {2},
          eid = {L39},
        pages = {L39},
          doi = {10.1088/2041-8205/758/2/L39},
archivePrefix = {arXiv},
       eprint = {1209.5393},
 primaryClass = {astro-ph.CO},
       adsurl = {https://ui.adsabs.harvard.edu/abs/2012ApJ...758L..39T}
}

@ARTICLE{2025arXiv250201024C,
       author = {{Chen}, Nianyi and {Di Matteo}, Tiziana and {Zhou}, Yihao and {Kelley}, Luke Zoltan and {Blecha}, Laura and {Ni}, Yueying and {Bird}, Simeon and {Yang}, Yanhui and {Croft}, Rupert},
        title = "{The Gravitational Wave Background from Massive Black Holes in the ASTRID Simulation}",
      journal = {arXiv e-prints},
         year = 2025,
        month = feb,
          eid = {arXiv:2502.01024},
        pages = {arXiv:2502.01024},
          doi = {10.48550/arXiv.2502.01024},
archivePrefix = {arXiv},
       eprint = {2502.01024},
 primaryClass = {astro-ph.GA},
       adsurl = {https://ui.adsabs.harvard.edu/abs/2025arXiv250201024C}
}

@article{Agarwal:2025cag,
    author = "Agarwal, Nikita and others",
    title = "{The NANOGrav 15 yr Data Set: Targeted Searches for Supermassive Black Hole Binaries}",
    eprint = "2508.16534",
    archivePrefix = "arXiv",
    primaryClass = "astro-ph.HE",
    month = "8",
    year = "2025"
}
\bibliographystyle{aasjournal}

\end{sloppypar}
\end{document}